\RequirePackage{fix-cm}
\documentclass[]{rsproca_new}

\usepackage{amssymb}
\usepackage{latexsym}
\usepackage{url}

\usepackage{amsmath}
\usepackage{subfigure}
\usepackage{epsfig}
\usepackage{siunitx}

\usepackage{color}
\usepackage[normalem]{ulem}

\usepackage{hyperref}   %

\makeatletter %

\def\input@path{{01_sections/}{02_figures/}}

\newcommand{\secref}[1]{Section~\ref{#1}}%
\newcommand{\figref}[1]{Figure~\ref{#1}}%

\newcommand\DeclareBoldMathCommand[2]{%
  \protected@edef\@tempb{%
    \noexpand\DeclareRobustCommand{\csname #1\endcsname}{\boldsymbol{\ensuremath{#2}}}}
  \@tempb}
\@tfor\AM@letter:=abcdefghijklmnopqrstuvwxyz%
\do{\DeclareBoldMathCommand{v\AM@letter}{\AM@letter}}
\@tfor\AM@letter:=ABCDEFGHIJKLMNOPQRSTUVWXYZ%
\do{\DeclareBoldMathCommand{v\AM@letter}{\AM@letter}}
\DeclareBoldMathCommand{vnull}{0}
\DeclareBoldMathCommand{vone}{1}
\DeclareBoldMathCommand{valpha}{\alpha}
\DeclareBoldMathCommand{vbeta}{\beta}
\DeclareBoldMathCommand{vgamma}{\gamma}
\DeclareBoldMathCommand{vdelta}{\delta}
\DeclareBoldMathCommand{vepsilon}{\epsilon}
\DeclareBoldMathCommand{vvarepsilon}{\varepsilon}
\DeclareBoldMathCommand{vzeta}{\zeta}
\DeclareBoldMathCommand{veta}{\eta}
\DeclareBoldMathCommand{vtheta}{\theta}
\DeclareBoldMathCommand{vvartheta}{\vartheta}
\DeclareBoldMathCommand{viota}{\iota}
\DeclareBoldMathCommand{vkappa}{\kappa}
\DeclareBoldMathCommand{vlambda}{\lambda}
\DeclareBoldMathCommand{vmu}{\mu}
\DeclareBoldMathCommand{vnu}{\nu}
\DeclareBoldMathCommand{vpi}{\pi}
\DeclareBoldMathCommand{vxi}{\xi}
\DeclareBoldMathCommand{vrho}{\varrho}
\DeclareBoldMathCommand{vsigma}{\sigma}
\DeclareBoldMathCommand{vtau}{\tau}
\DeclareBoldMathCommand{vupsilon}{\upsilon}
\DeclareBoldMathCommand{vphi}{\varphi}
\DeclareBoldMathCommand{vchi}{\chi}
\DeclareBoldMathCommand{vpsi}{\psi}
\DeclareBoldMathCommand{vomega}{\omega}
\DeclareBoldMathCommand{vGamma}{\Gamma}
\DeclareBoldMathCommand{vDelta}{\Delta}
\DeclareBoldMathCommand{vTheta}{\Theta}
\DeclareBoldMathCommand{vLambda}{\Lambda}
\DeclareBoldMathCommand{vXi}{\Xi}
\DeclareBoldMathCommand{vPi}{\Pi}
\DeclareBoldMathCommand{vSigma}{\Sigma}
\DeclareBoldMathCommand{vUpsilon}{\Upsilon}
\DeclareBoldMathCommand{vPhi}{\Phi}
\DeclareBoldMathCommand{vPsi}{\Psi}
\DeclareBoldMathCommand{vOmega}{\Omega}

\newcommand\DeclareDiscreteBoldMathCommand[2]{%
  \protected@edef\@tempc{%
    \noexpand\DeclareRobustCommand{\csname #1\endcsname}{\boldsymbol{\mathrm{#2}}}}
  \@tempc}
\@tfor\AM@letter:=abcdefghijklmnopqrstuvwxyz%
\do{\DeclareDiscreteBoldMathCommand{vd\AM@letter}{\AM@letter}}
\@tfor\AM@letter:=ABCDEFGHIJKLMNOPQRSTUVWXYZ%
\do{\DeclareDiscreteBoldMathCommand{vd\AM@letter}{\AM@letter}}
\DeclareDiscreteBoldMathCommand{vdnull}{0}
\DeclareDiscreteBoldMathCommand{vdone}{1}
\DeclareDiscreteBoldMathCommand{vdalpha}{\alpha}
\DeclareDiscreteBoldMathCommand{vdbeta}{\beta}
\DeclareDiscreteBoldMathCommand{vdgamma}{\gamma}
\DeclareDiscreteBoldMathCommand{vddelta}{\delta}
\DeclareDiscreteBoldMathCommand{vdepsilon}{\epsilon}
\DeclareDiscreteBoldMathCommand{vdvarepsilon}{\varepsilon}
\DeclareDiscreteBoldMathCommand{vdzeta}{\zeta}
\DeclareDiscreteBoldMathCommand{vdeta}{\eta}
\DeclareDiscreteBoldMathCommand{vdtheta}{\theta}
\DeclareDiscreteBoldMathCommand{vdvartheta}{\vartheta}
\DeclareDiscreteBoldMathCommand{vdiota}{\iota}
\DeclareDiscreteBoldMathCommand{vdkappa}{\kappa}
\DeclareDiscreteBoldMathCommand{vdlambda}{\lambda}
\DeclareDiscreteBoldMathCommand{vdmu}{\mu}
\DeclareDiscreteBoldMathCommand{vdnu}{\nu}
\DeclareDiscreteBoldMathCommand{vdpi}{\pi}
\DeclareDiscreteBoldMathCommand{vdxi}{\xi}
\DeclareDiscreteBoldMathCommand{vdrho}{\varrho}
\DeclareDiscreteBoldMathCommand{vdsigma}{\sigma}
\DeclareDiscreteBoldMathCommand{vdtau}{\tau}
\DeclareDiscreteBoldMathCommand{vdupsilon}{\upsilon}
\DeclareDiscreteBoldMathCommand{vdphi}{\varphi}
\DeclareDiscreteBoldMathCommand{vdchi}{\chi}
\DeclareDiscreteBoldMathCommand{vdpsi}{\psi}
\DeclareDiscreteBoldMathCommand{vdomega}{\omega}
\DeclareDiscreteBoldMathCommand{vdGamma}{\Gamma}
\DeclareDiscreteBoldMathCommand{vdDelta}{\Delta}
\DeclareDiscreteBoldMathCommand{vdTheta}{\Theta}
\DeclareDiscreteBoldMathCommand{vdLambda}{\Lambda}
\DeclareDiscreteBoldMathCommand{vdXi}{\Xi}
\DeclareDiscreteBoldMathCommand{vdPi}{\Pi}
\DeclareDiscreteBoldMathCommand{vdSigma}{\Sigma}
\DeclareDiscreteBoldMathCommand{vdUpsilon}{\Upsilon}
\DeclareDiscreteBoldMathCommand{vdPhi}{\Phi}
\DeclareDiscreteBoldMathCommand{vdPsi}{\Psi}
\DeclareDiscreteBoldMathCommand{vdOmega}{\Omega}

\providecommand*{\dd}{%
  \@ifnextchar^{\@dd}{\@dd^{}}}
\def\@dd^#1{%
  \mathop{\mathrm{\mathstrut d}}%
  \nolimits^{#1}\dd@gobblespace}
\def\dd@gobblespace{%
  \futurelet\diffarg\dd@opspace}
\def\dd@opspace{%
  \let\dd@space\!%
  \ifx\diffarg(%
\let\dd@space\relax%
\else%
\ifx\diffarg[%
\let\dd@space\relax%
\else%
\ifx\diffarg\{%
\let\dd@space\relax%
\fi%
\fi%
\fi%
\dd@space}

\newcommand{\Frac}{%
  \@ifnextchar[%
  {\Frac@i}
  {\Frac@ii}}
\newcommand{\Frac@i}{}
\def\Frac@i[#1]#2#3{%
  \genfrac{}{}{#1}{}{\displaystyle{#2}}{\displaystyle{#3}}}
\newcommand{\Frac@ii}[2]{\frac{\displaystyle{#1}}{\displaystyle{#2}}}
      \newcommand{\diff@diffspace}{\,}
\newcommand{\diff@mathfrac}[2]{\frac{#1}{#2}}
\newcommand{\diff@mathFrac}[2]{\Frac{#1}{#2}}
\newcommand{\diff@textfrac}[2]{%
  \bgroup #1\egroup\mkern-1mu/\mkern-1mu\bgroup #2\egroup}
\newcommand{\diff}{%
  \global\let\diff@diffop\dd
  \global\let\diff@frac\diff@mathfrac
  \@ifnextchar[%
  {\diff@i}
  {\diff@ii}}
\newcommand{\Diff}{%
  \global\let\diff@diffop\dd
  \global\let\diff@frac\diff@mathFrac
  \@ifnextchar[%
  {\diff@i}
  {\diff@ii}}
\newcommand{\tdiff}{%
  \global\let\diff@diffop\dd
  \global\let\diff@frac\diff@textfrac
  \@ifnextchar[%
  {\diff@i}
  {\diff@ii}}
\newcommand{\pdiff}{%
  \global\let\diff@diffop\partial
  \global\let\diff@frac\diff@mathfrac
  \@ifnextchar[%
  {\diff@i}
  {\diff@ii}}
\newcommand{\Pdiff}{%
  \global\let\diff@diffop\partial
  \global\let\diff@frac\diff@mathFrac
  \@ifnextchar[%
  {\diff@i}
  {\diff@ii}}
\newcommand{\tpdiff}{%
  \global\let\diff@diffop\partial
  \global\let\diff@frac\diff@textfrac
  \@ifnextchar[%
  {\diff@i}
  {\diff@ii}}

\newcommand*{\diff@i}{}
\def\diff@i[#1]#2#3{\eval{\diff@ii{#2}{#3}}_{#1}}

\newcommand*{\diff@ii}[2]{%
  \begingroup
  \toks0={}\count0=0
  \diff@degree #2\diff@degree
  \diff@frac{\diff@diffop\ifnum\count0>1^{\the\count0}\fi\diff@diffspace#1}%
  {\the\toks0}%
  \endgroup}

\newcommand*{\diff@degree}[1]{%
  \ifx #1\diff@degree \expandafter\diff@stopd
  \else \expandafter\diff@addd \fi #1^1$#1\diff@addd}
\newcommand{\diff@stopd}{}
\def\diff@stopd #1\diff@addd{}
\newcommand*{\diff@addd}{}
\def\diff@addd #1^#2#3$#4\diff@addd{%
  \advance\count0 #2
  \toks0=\expandafter{\the\toks0%
    {\diff@diffop\diff@diffspace #4}%
    \diff@diffspace}\diff@degree}

\def\rs#1{\@ifnextchar[%
  {\@rs{#1}}{\@@rs{#1}}}
\def\@rs#1[#2]#3{\mathinner{%
    \setbox\@ne\hbox{$\displaystyle{\vphantom{#3}}#1{#3}\m@th$}%
    \setbox\tw@\hbox{$\displaystyle{#3}#2\m@th$}%
    \hskip\wd\@ne\hskip-\wd\tw@\mathord{\hskip\wd\tw@\hskip-\wd\@ne%
      {\vphantom{#3}}#1{#3}#2}}}
\def\@@rs#1#2{\mathinner{%
    \setbox\@ne\hbox{$\displaystyle{\vphantom{#2}}#1{#2}\m@th$}%
    \hskip\wd\@ne\mathord{\hskip-\wd\@ne%
      {\vphantom{#2}}#1{#2}}}}

\newcommand{\MR}{\mathalpha{\mathbb{R}}}

\newcommand*{\norm}[1]{\mathinner{\Vert#1\Vert}}

\definecolor{notecolor}{cmyk}{0,1,1,.2}
\newcommand*\AM@notesname{Notes}

\makeatother

\graphicspath{{02_figures/}}

\jname{rspa}
\Journal{Proc R Soc A\ }

\begin{document}

\title{A Novel Modeling and Simulation Approach for the Hindered Mobility of Charged Particles in Biological Hydrogels}

\author{Maximilian J.~Grill$^{1}$, Jonas F.~Eichinger$^{1}$, Jonas Koban$^{1}$, Christoph Meier$^{1}$, Oliver Lieleg$^{2}$ and Wolfgang A.~Wall$^{1}$}

\address{$^{1}$ Institute for Computational Mechanics, Technical University of Munich, Germany\\
$^{2}$ Munich School of Bioengineering, Technical University of Munich, Germany}

\subject{computational biophysics, modeling and simulation, biomaterials}


\corres{Maximilian J.~Grill\\
\email{grill@lnm.mw.tum.de}}

\begin{abstract}
This article presents a novel computational model to study the selective filtering of biological hydrogels due to the surface charge and size of diffusing particles.
It is the first model that includes the random 3D fiber orientation and connectivity of the biopolymer network and that accounts for elastic deformations of the fibers by means of beam theory.
As a key component of the model, novel formulations are proposed both for the electrostatic and repulsive steric interactions between a spherical particle and a beam.
In addition to providing a thorough validation of the model, the presented computational studies yield new insights into the underlying mechanisms of hindered particle mobility, especially regarding the influence of the aforementioned aspects that are unique to this model.
It is found that the precise distribution of fiber and thus charge agglomerations in the network have a crucial influence on the mobility of oppositely charged particles and gives rise to distinct motion patterns.
Considering the high practical significance for instance with respect to targeted drug release or infection defense, the provided proof of concept motivates further advances of the model toward a truly predictive computational tool that allows a case- and patient-specific assessment for real (biological) systems.
\end{abstract}

\begin{fmtext}

\end{fmtext}

\maketitle

\section{Introduction}
\label{sec::particle_mobility_introduction}
The remarkable ability of hydrogel forming biopolymer networks to control the mobility of different kinds of diffusing molecules and particles individually is of crucial importance for numerous functions of the human body.
See~\cite{Witten2017} for a recent review of this topic.
On the one hand, this selective permeability gives rise to the protection of the organism against pathogens such as viruses that are effectively hindered from invading and traversing the organism.
On the other hand, it ensures the effective transport of a broad variety of substances that are useful and important for the organism.
Examples of such biological hydrogels include mucus, the extracellular matrix (ECM), intracellular biopolymer networks (comprising, e.g., neurofilaments or actin), the vitreous humor and the matrix of biofilms, and can thus be found throughout the entire human body and in countless other places in nature.
The high practical relevance of this fundamental topic extends over multiple fields ranging from medical diagnosis to the therapy of body malfunctions and targeted drug release.
This creates yet unimagined possibilities in technical applications such as filters used in chemical, mechanical or medical engineering.

A large number of experimental studies have investigated the origin of this selective filtering.
Meanwhile, there is strong evidence that -- besides the most obvious mechanism to filter by size -- the surface properties of the particles also plays an important role (see e.g.,~\cite{Witten2017,Bray1984,Dowd1999,Dellian2000,Olmsted2001,Lai2007,Lieleg2009,Lieleg2010,Colwell2010,Schuster2013,Xu2013,Arends2013,Zhang2015,Kasdorf2015,Abdulkarim2015,Arends2017}).
This new paradigm thus suggests filtering not (only) by size but also through a combination of other particle properties such as charge, hydrophobicity or binding affinity and can thus be referred to as interaction filtering.
Despite considerable scientific effort in this field, many aspects concerning the underlying mechanisms and specific conditions remain unknown.
To a large extent, this can be explained by the following three factors.
First, there is a great complexity in the many different biological systems both in number and diversity of components, e.g., with respect to their molecular architecture.
Second, there are big challenges and limitations with respect to experimental preparation and measurement techniques when it comes to the required high spatial and temporal resolution, especially over considerable time spans of several seconds.
And third, there is a fundamental lack of in-depth understanding of physical and chemical interactions on the molecular scale.
This lack of detailed, fundamental microscopic understanding prevents any reliable prediction of the diffusive mobility of molecules and particles in other than the few particle-hydrogel systems already studied in vitro.
At this point, the development of accurate and efficient computational models, which are capable of resolving small scales and covering large spans in space and time, is expected to substantially contribute to scientific progress in this field.

Compared to the large body of literature reporting on experimental work in this area, relatively few computational studies have been published so far.
Most of the early and also some of the recent work focused solely on the excluded volume effect of the fibers, i.e., the repulsive steric interactions with the network fibers that hindered the free diffusion of a (hard) spherical particle (see e.g.,~\cite{Johansson1993,Saxton1994,Netz1997,Pei2009,Stylianopoulos2010fiberorientation,Kamerlin2016}).
In his Monte Carlo simulations, Saxton \cite{Saxton1996} was the first to include and study other than steric interactions in the form of a binding model.
Since the recognition of the dominant role of electrostatic and possibly other types of molecular interactions as outlined above, several computational studies have included electrostatic effects and confirmed the trends observed in experiments and shed light on the underlying mechanisms \cite{Zhou2009,Stylianopoulos2010electrostatics,Miyata2012,Zhang2015,Hansing2016,Hansing2018}.
Particularly the recent works published by Hansing et al.~\cite{Hansing2016} and Hansing and Netz \cite{Hansing2018} were very successful in analyzing the influences of particle size, fiber volume fraction, particle charge, and the comparison of oppositely vs.~similarly charged particles and networks.
They also found good agreement of the simulation results with several sets of experimental data, which confirms the validity of such modeling approach.

The computational model proposed in this work aims to improve especially the modeling of the fiber network, which has been modeled in a very simplified manner in all previous studies.
Either a square array of straight and parallel rigid fibers oriented along one spatial dimension \cite{Stylianopoulos2010electrostatics} or a cubic lattice consisting of either linearly aligned hard spheres \cite{Miyata2012} or consisting of straight rigid fibers \cite{Zhang2015,Hansing2016} has been assumed.
Zhou and Chen~\cite{Zhou2009} likewise applied a cubic lattice consisting of beads placed at the vertices and connected by linear spring elements.
Hansing and Netz \cite{Hansing2018} were the first to break the strong geometrical order in all of these models, they only allowed, however, for a variation of the spacing of the still straight and infinitely long rigid fibers with a mutually orthogonal orientation.
Yet they found a significant and fundamental difference in the trapping mechanism of ordered and disordered fiber lattices, which can be attributed to locally denser regions of the network in the case of attractive particle-network interactions.
This is a strong motivation to work toward a more realistic modeling of the fiber network as a crucial part of biological hydrogels as will be outlined in the following paragraphs.

In our approach, the individual fibers will be modeled by the geometrically exact 3D beam theory, thus allowing for arbitrarily curved initial shapes of the fibers and possibly large deformations, including all six modes of axial strain, (2x) shear, torsion and (2x) bending.
In addition, the initial spatial distribution, orientation and interconnection of fibers will be modeled as a random 3D Voronoi network, thus mimicking several important geometrical features of real biopolymer networks such as a random, spatially variable mesh size distribution, a random fiber length distribution, random mutual orientations of the fibers, and arbitrary connectivity between the fibers.
Several of these attributes are expected to play an important role when it comes to both purely steric interactions and the combination with electrostatic interactions and will be studied in \secref{sec::particle_mobility_results}.
A similar approach to network generation based on Voronoi tessellation has previously been applied in a number of publications, e.g., to study cell-cell communication in a 2D network of linearly elastic springs \cite{Humphries2017}.

Applying such a sophisticated model to the individual fibers and the network they constitute comes at the cost of an increased complexity and size of the system of equations to be solved in the simulations.
However, the same modeling strategy based on geometrically exact beam finite elements describing the biopolymer fibers has previously been applied to model the Brownian dynamics of individual semiflexible filaments \cite{Cyron2010,Slepukhin2019} and has been proven to be accurate and also efficient enough to study large-scale problems such as the process of self-assembly of (different morphologies of) transiently cross-linked biopolymer networks \cite{Cyron2013phasediagram,Cyron2013a} as well as their (high- and low-frequency) rheology \cite{Mueller2014rheology}.
On the one hand, this confirms that the novel modeling approach is suited to study the hindered mobility of particles in hydrogels and on the other hand, this already outlines the long-term opportunities.
Using the self-assembled network configurations, e.g., for an actin bundle network, can readily replace the Voronoi-type network applied as a first step in the present study.
Moreover, the dynamics of the network, including the reorganization of the transient cross-links could be included and would be highly interesting since there is experimental evidence that particles larger than the mesh size can still diffuse through the network by locally breaking inter-fiber links \cite{Lai2007}.
This has also been confirmed to be an effective transport mechanism in a first computational model \cite{Goodrich2018}.
Another example for the dynamics of hydrogel networks is given by the self-renewal process of a mucus layer \cite{Witten2017,Marczynski2018}.

In general, the mathematical description of the problem and therefore also the numerical methods required to solve it as well as the code framework substantially differs from the ones used by the previous computational studies listed above.
This is an inevitable consequence of the aforementioned modeling of individual biopolymer fibers as elastically deformable, geometrically exact beams.
In the present model, a set of well-established numerical formulations for beams \cite{Meier2017c,Meier2017b} is combined with novel beam-sphere interaction models, which are derived as a special case of the more general approach to fiber-fiber interactions recently proposed in \cite{GrillSSIP}.
The latter is the first 3D beam-beam interaction model for molecular interactions between curved slender fibers undergoing large deformations and is thus an important prerequisite for the beam-sphere interaction formulations to be applied in this article.

The remainder of this article is structured as follows.
\secref{sec::particle_mobility_model_setup} presents all aspects of the computational model and the required numerical methods.
After the presentation and discussion of the results in \secref{sec::particle_mobility_results}, this article will be concluded with a summary of the findings and an outlook to promising aspects of future research in \secref{sec::particle_mobility_conclusion_outlook}.

\section{Computational model and methods}
\label{sec::particle_mobility_model_setup}
Given the broad variety of biological hydrogels mentioned above, we chose the ECM gel(s) used in the comprehensive experimental studies of particle mobility by Lieleg et al.~\cite{Lieleg2009} and Arends et al.~\cite{Arends2013} to serve as the main reference for the specific setup and parametrization of the versatile computational model.
For these gels, an additional, subsequent characterization in terms of their biophysical properties has been conducted \cite{Arends2015PLOS}, which shall prove useful in the following.
To reduce the complexity of this multi-component biological system, the computational model considers only the following three key components, which are thought to mimic the crucial influences of the studied problem: a fiber network, a diffusing particle, and their mutual interactions.
All three model components will be considered individually in the following and selected further aspects of the simulation setup such as boundary conditions and post-processing of the results will be discussed.
The software package used for the simulations in this article is the parallel, multi-physics, in-house research code BACI~\cite{BACI2020}.

\subsection{Overall approach}
Following up on the concept of modeling biopolymer fibers as elastically deformable, geometrically exact beams as described above, the overall approach follows the one commonly used in nonlinear finite element frameworks for structural dynamics.
In short, this approach consists of the following steps:
According to the principle of virtual work, the weak form of the mechanical balance equations is derived and subsequently discretized in space and time.
Given a proper set of initial and boundary conditions, an implicit load/time stepping scheme is applied and in every step the solution of the resulting discrete system of nonlinear equations is found iteratively by means of Newton's method.
Refer to Ref.~\cite{GrillSSIP} for a discussion of selected aspects of the applied algorithms and libraries.

\subsection{Biopolymer fiber network}
\label{sec::particle_mobility_model_setup_fiber_network}
\begin{figure}[htpb]%
  \centering
  \subfigure[]{
    \includegraphics[width=\textwidth]{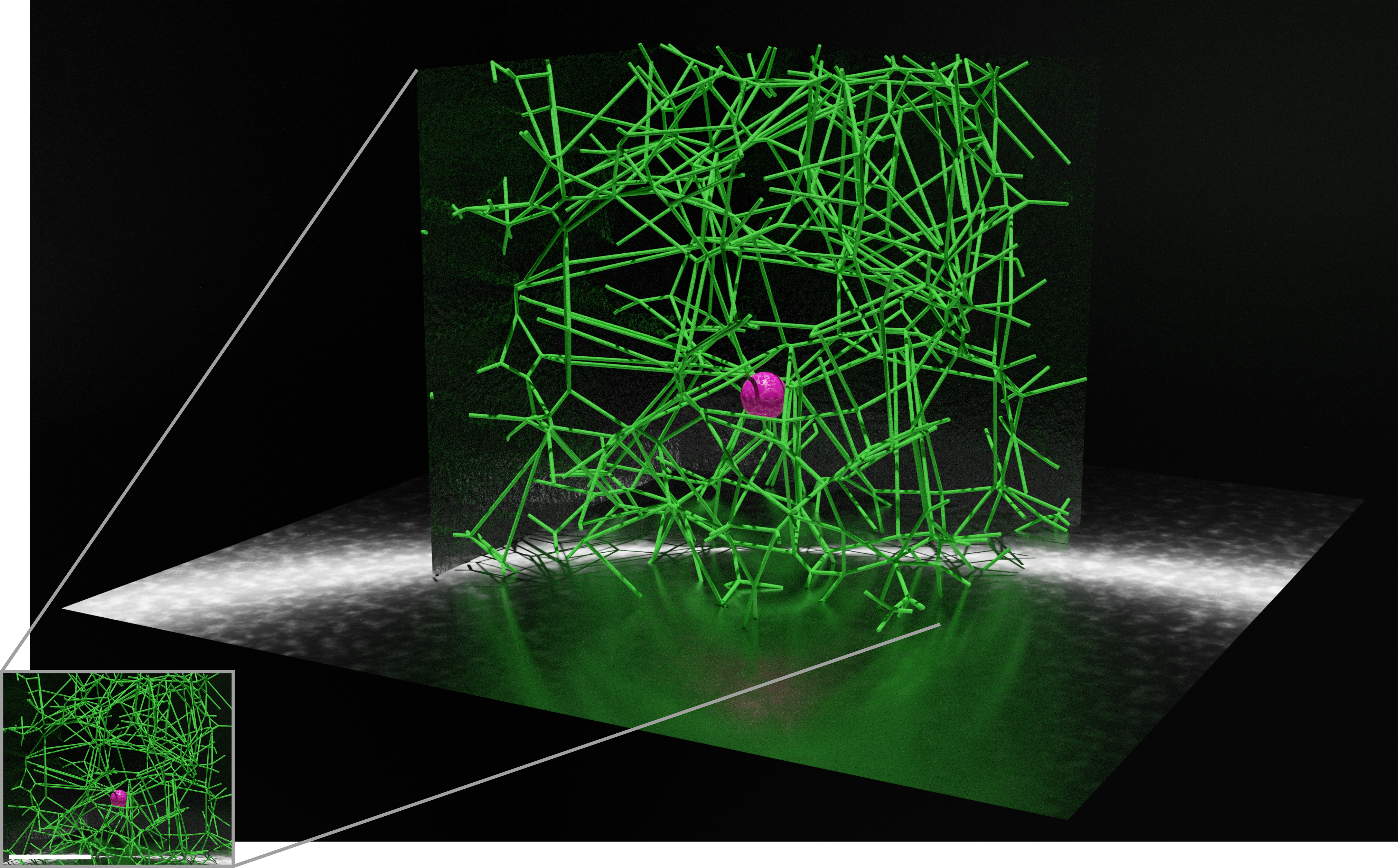}
    \label{fig::particle_mobility_network_60VP}
  }
  \subfigure[]{
    \includegraphics[width=\textwidth]{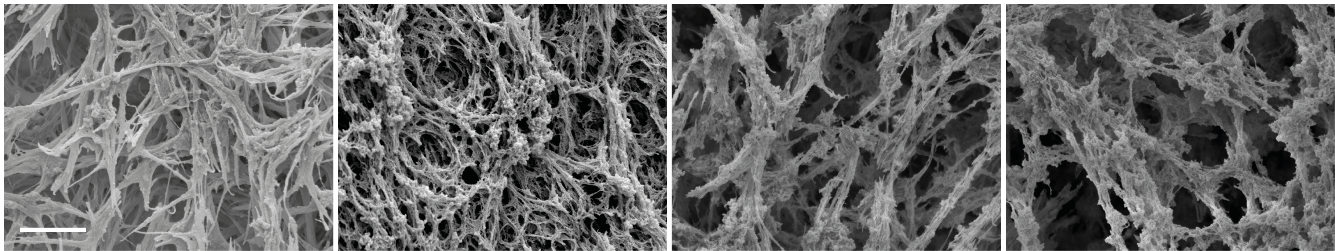}
    \label{fig::basal_lamina_gel_SEM_images_5mu}
  }
  \subfigure[]{
    \includegraphics[width=\textwidth]{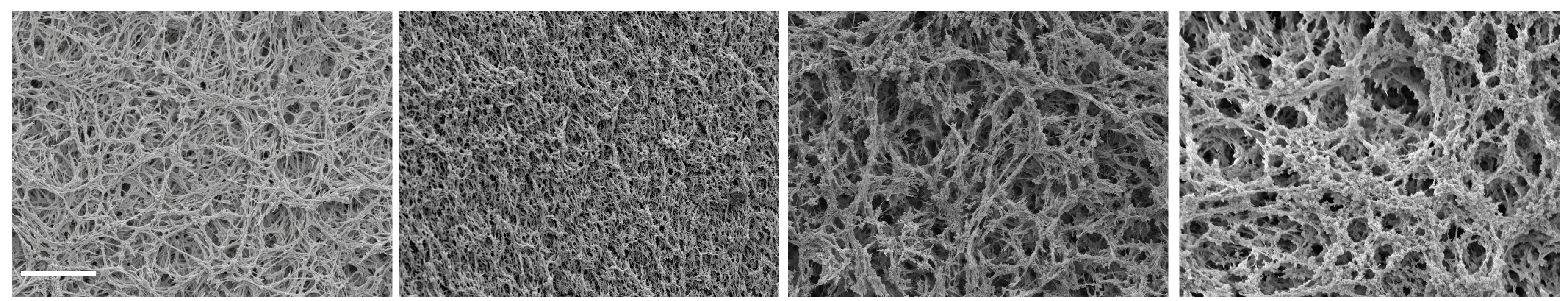}
    \label{fig::basal_lamina_gel_SEM_images_25mu}
  }
  \caption{Network architectures: (a) The result of a random, periodic Voronoi tessellation, which is used as the initial, stress-free configuration in the computational model (box size is $10\times10\times10 \, \SI{}{\micro\meter}$, rendered using Blender \cite{Blender2_80}).
  The inset shows the scaled system (scale bar indicates $\SI{5}{\micro\meter}$) used to ease the comparison with the other images.
  (b)--(c) SEM images of four different basal lamina gels used in in vitro experiments (reprinted from \cite{Arends2015PLOS}, scale bars indicate (b) $\SI{5}{\micro\meter}$ and (c) $\SI{25}{\micro\meter}$).}
  \label{fig::particle_mobility_setup_networks}
\end{figure}

As outlined above, the initial, stress-free configuration of the fiber network is the result of a 3D Voronoi tessellation of the cubic simulation box (size $10\times10\times10 \, \SI{}{\micro\meter}$) generated via the open source library voro++ \cite{Rycroft2009}.
\figref{fig::particle_mobility_network_60VP} shows an example of the resulting network architecture.
The main input of this preprocessing step are the randomly chosen locations of a number $n_\text{VP}$ of so-called Voronoi points%
\footnote{Note that the original term Voronoi \textit{particles} from \cite{Rycroft2009} is not used here to avoid confusion with the diffusing particle(s).
}.
The output are the vertices and edges of a random, irregular, polygonal network that are used to define the position and orientation of the initially straight beam segments as well as their interconnections.
For Voronoi-based tessellation, the connectivity number, i.e., the number of fibers branching off at each junction point, is 4, which agrees well with values of 3 to 4 reported for ECM gels \cite{Yurchenco1987}.
In the future, this parameter could also be adapted by randomly removing or adding fibers to match other hydrogel architectures.
At each junction, the beam endpoint positions and rotations are coupled, which corresponds to a model assumption based on the expected mechanical behavior of the chemical binding between the fibers, and could be once again adapted e.g., to hinged connections in a straightforward manner.
Also, the binding and unbinding of such connections at given rates could be included by following the approach described e.g.,~in \cite{Mueller2014rheology}.
To be able to use the simulation box as a representative volume element (RVE) with periodic boundary conditions (see \secref{sec::particle_mobility_model_setup_boundary_conditions} for details), the Voronoi tessellation and thus the resulting network geometry is chosen to be periodic in space.
Altogether, a simple visual comparison with electron microscopy images of real biological hydrogels in~\figref{fig::particle_mobility_setup_networks} reveals a high similarity both in terms of the random, irregular, polygonal structure as well as its characteristic properties, such as the distribution of pore sizes, fiber segment length and connectivity.

The individual fibers are modeled by (geometrically exact) 3D beam theory, assuming undeformable cross-sections of circular shape and a hyperelastic material law.
Specifically, we apply the Simo-Reissner beam theory \cite{reissner1981,simo1985,simo1986}, which accounts for 6 deformation modes of axial tension, (2x) shear, torsion and (2x) bending.
Further specifications such as the dimensional and constitutive parameters are chosen to mimic collagen I as the key constituent of the targeted class of ECM hydrogels, however, all these parameters can be easily adapted to study their influences or to model other fiber species.
In the present study, the cross-section diameter is set to~$D_\text{f} = \SI{75}{\nano\meter}$, Poisson's ratio is set to~$\nu = 0.3$, and Young's modulus is varied from as low as~$E=\SI{0.1}{\mega\pascal}$ up to the theoretical limit of rigid fibers to study the influence of the fiber stiffness.
Note that experimental measurements for collagen I suggest values in the wide range of~$E=\SI{1}{\mega\pascal}-\SI{1}{\giga\pascal}$ \cite{Jansen2018,Rijt2006}, which are covered in this work as well.
Based on the results in~\secref{sec::particle_mobility_results_varying_stiffness}, it will turn out that the influence of the fiber stiffness on the particle mobility in the problem setup considered here is negligible for the realistic range of values for Young's modulus~$E$, and a noticeable difference in results can only be observed below a threshold value of~$E^\ast=\SI{1}{\mega\pascal}$.
It can therefore be assumed that the specific values for the fiber material are of minor importance and that the deformation of fibers will only become important if the dynamics of the network reorganization will be included as outlined above, or if much thinner and softer, i.e., much more compliant fibers are considered e.g.,~in the context of a different kind of hydrogel (e.g.,~mucin or F-actin) or a dysregulation of fiber stiffness.

\begin{figure}[htb]%
  \centering
  \subfigure[]{
    \includegraphics[width=0.48\textwidth]{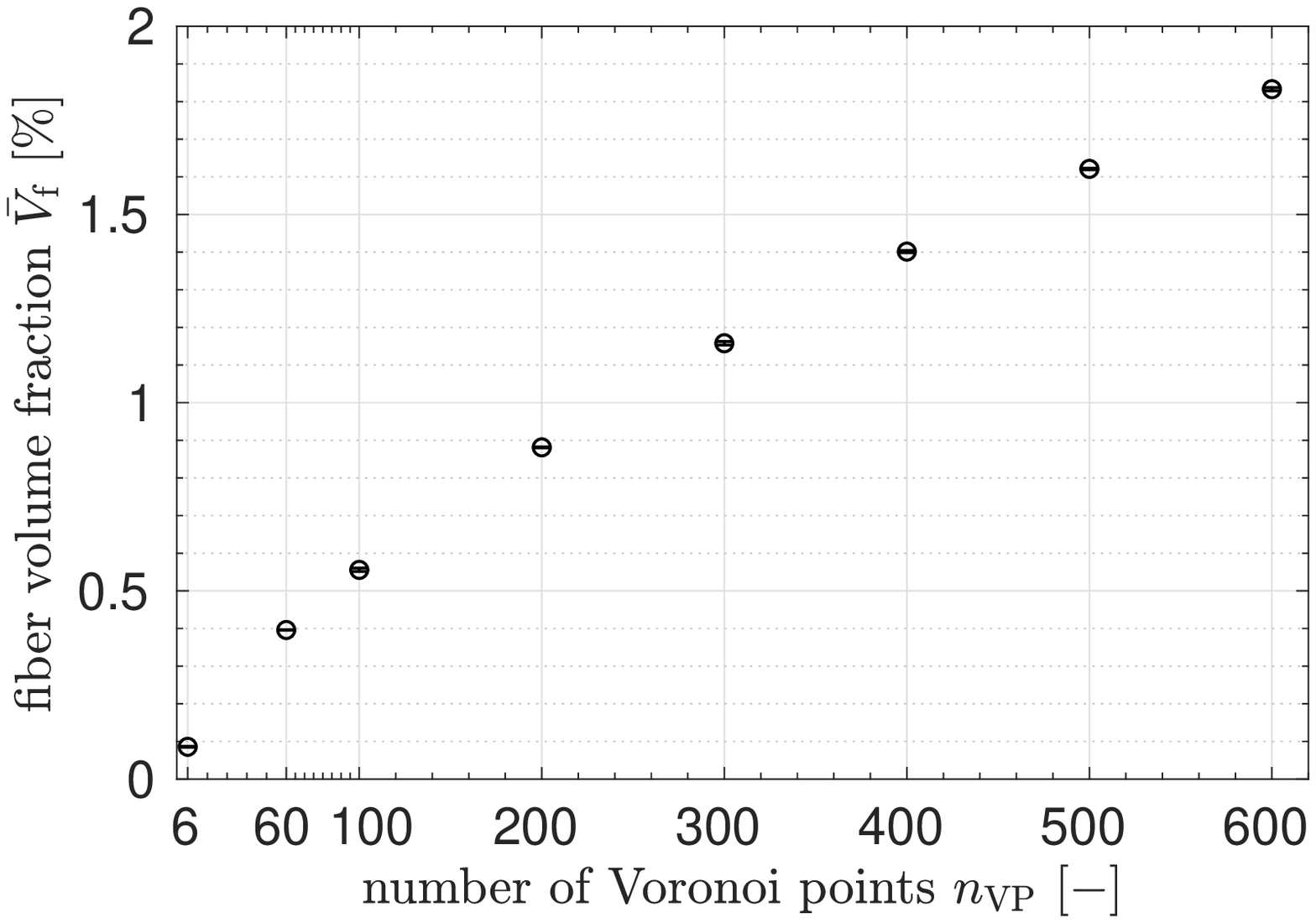}
    \label{fig::particle_mobility_fiber_volume_fraction_over_num_Voronoi_points}
  }
  \subfigure[]{
    \includegraphics[width=0.48\textwidth]{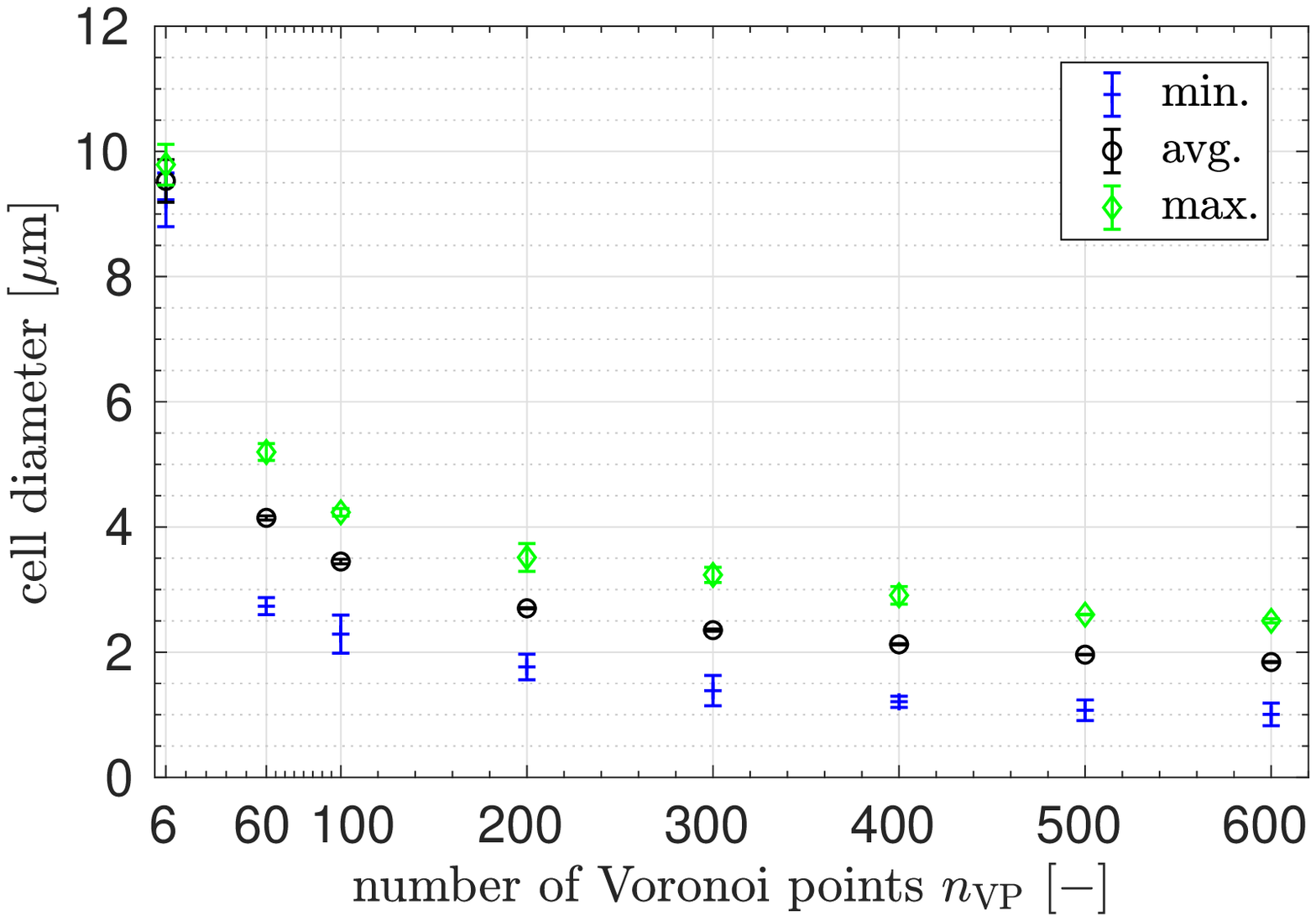}
    \label{fig::particle_mobility_cell_size_distribution_over_num_Voronoi_points}
  }
  \caption{(a) Mean and standard deviation of the fiber volume fraction (black circles with error bars, which are smaller than the symbol size).
          (b) Mean and standard deviation of the minimum cell diameter (blue pluses with error bars) / average cell diameter (black circles with error bars, again smaller than the symbol size) / maximum cell diameter (green diamonds with error bars), obtained for three random network geometries for each of the considered numbers of Voronoi points~$n_\text{VP}$.}
  \label{fig::particle_mobility_Voronoi_network_geometry_analysis}
\end{figure}

\begin{figure}[htpb]%
  \centering
  \subfigure[]{
    \includegraphics[width=\textwidth]{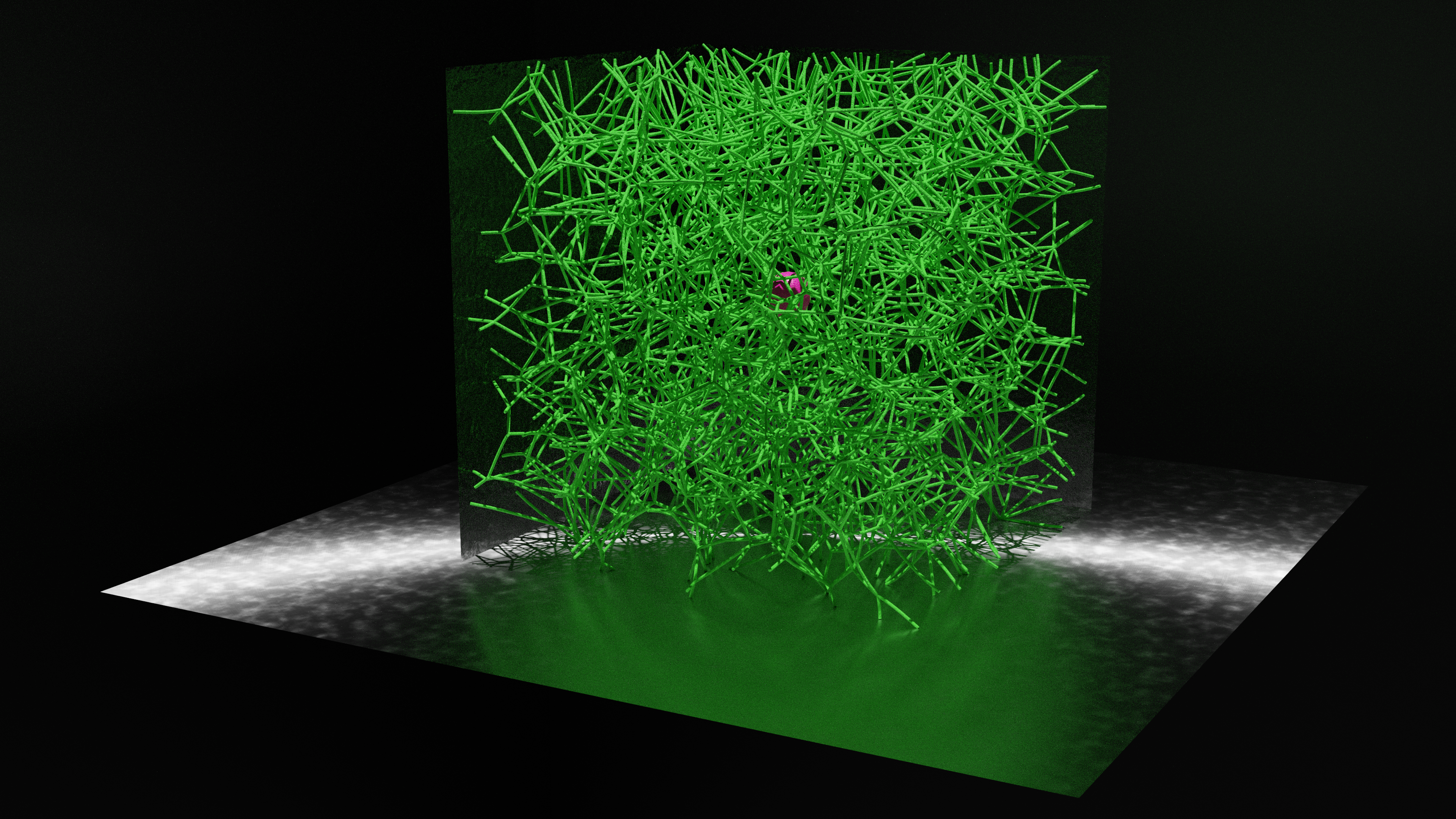}
    \label{fig::particle_mobility_network_600VP}
  }
  \subfigure[]{
    \includegraphics[width=0.45\textwidth]{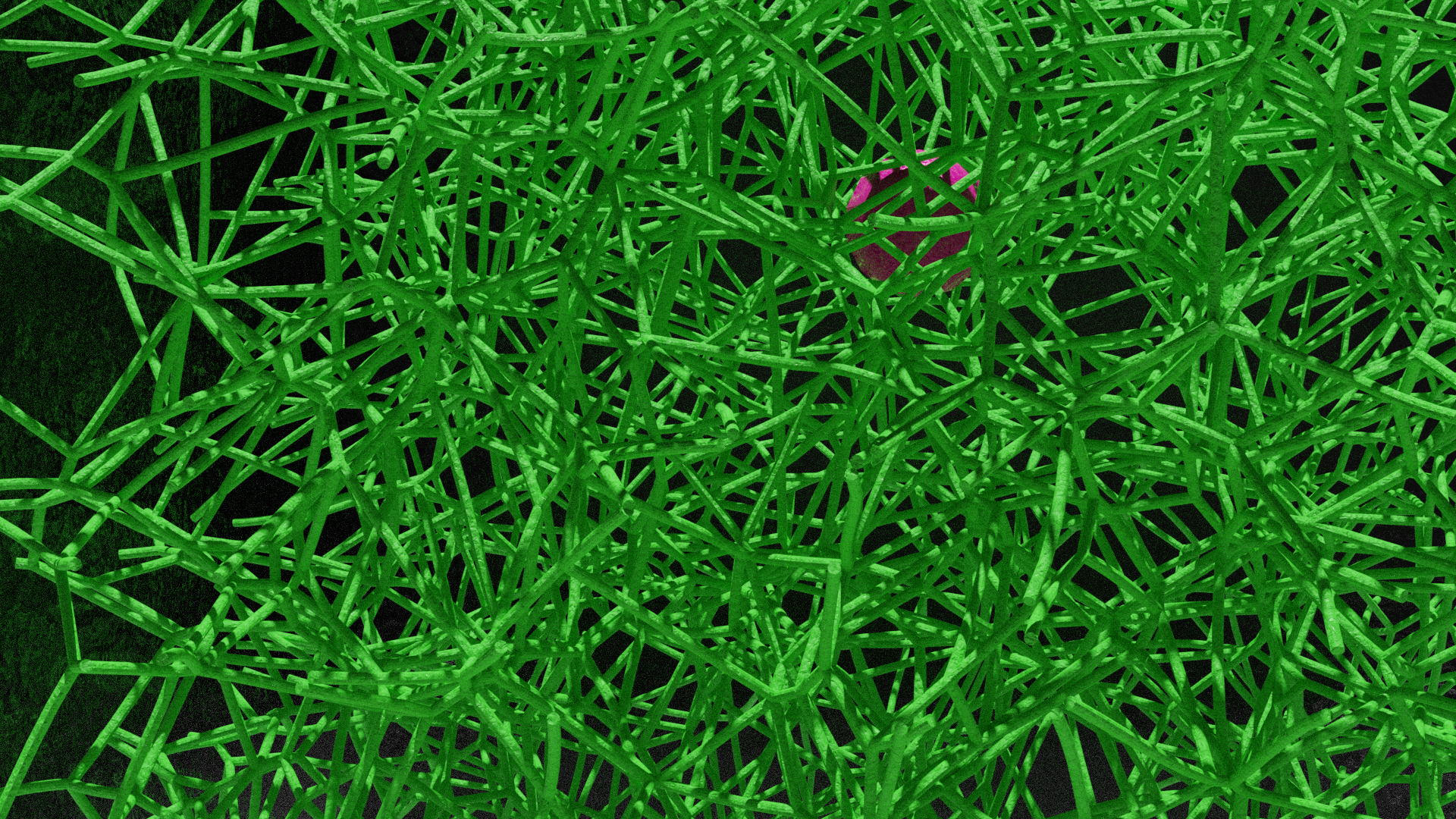}
    \label{fig::particle_mobility_network_600VP_detail}
  }
  \subfigure[]{
    \includegraphics[width=0.45\textwidth]{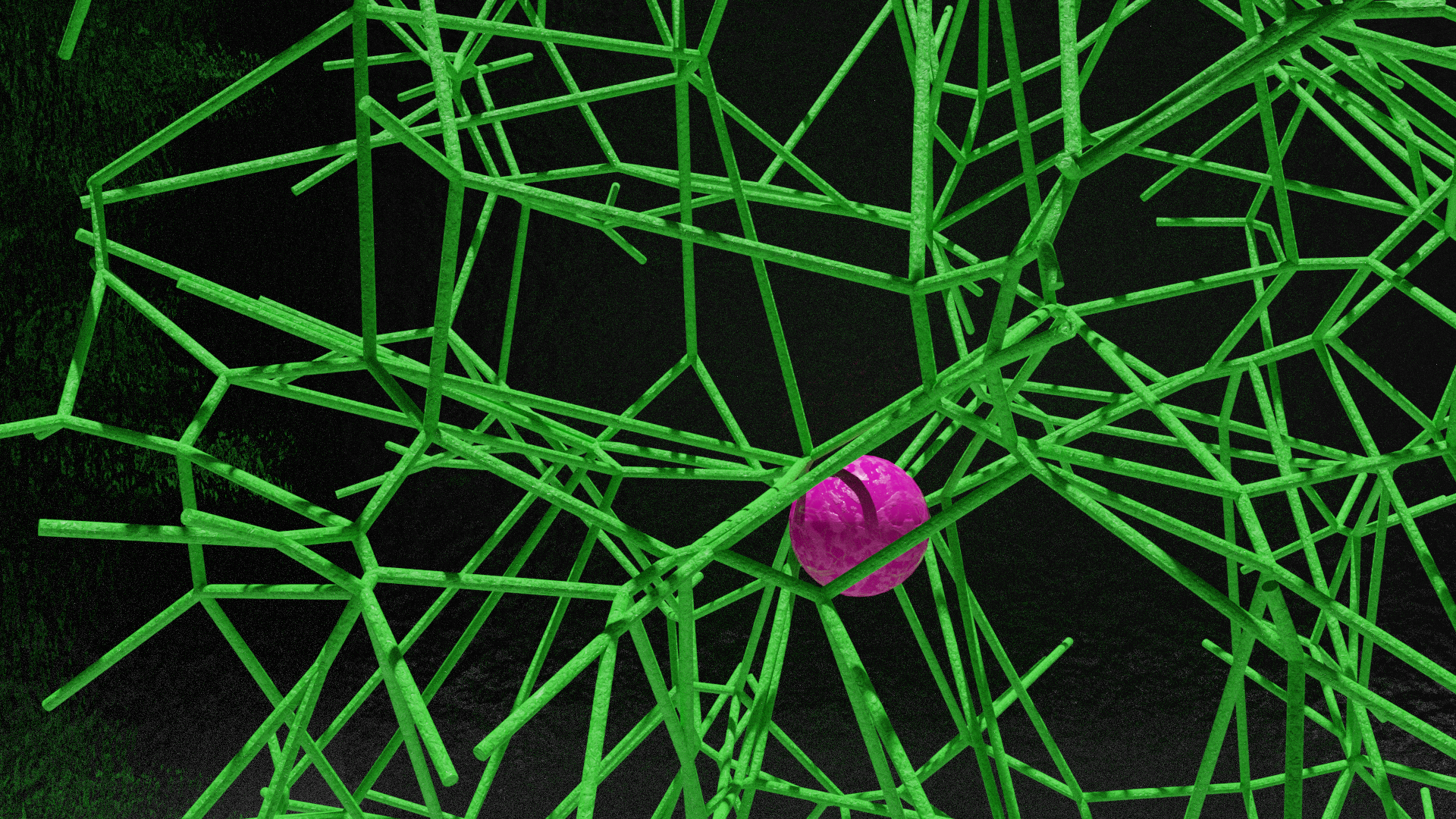}
    \label{fig::particle_mobility_network_60VP_detail}
  }
  \caption{(a) Image of an example network resulting for a second, very high value for the fiber volume fraction~$\bar V_\text{f} \approx 1.8\%$ considered in the simulations (box size is $10\times10\times10 \, \SI{}{\micro\meter}$, rendered using Blender \cite{Blender2_80}). Comparison of (b) a magnified part of this network with (c) a magnified part of the network with the default fiber volume fraction~$\bar V_\text{f} \approx 0.4\%$ (shown in \figref{fig::particle_mobility_network_60VP}).
  }
  \label{fig::particle_mobility_setup_networks_comparison_highfibervolfrac}
\end{figure}

In order to characterize the generated fiber networks, \figref{fig::particle_mobility_Voronoi_network_geometry_analysis} shows the resulting values of the mean and standard deviation of the fiber volume fraction~$\bar V_\text{f}$ and the minimum/average/maximum cell diameters%
\footnote{%
Here, the cell diameter for each of the $n_\text{VP}$ cells in the network has been computed as an approximated inscribed sphere of the irregular polyhedron based on the shortest distance of the Voronoi point to each of the cell edges.
}
as a function of the number of Voronoi points~$n_\text{VP}$.
For this statistical analysis of the random network geometries, a box size of $10\times10\times10 \, \SI{}{\micro\meter}$ and a fiber diameter~$D_\text{f} = \SI{75}{\nano\meter}$ as stated above is assumed.
A number of $n_\text{VP} = 60$ Voronoi points result in a fiber volume fraction~$\bar V_\text{f} \approx 0.4\%$ and a range of cell diameters of approx.~$2.7-\SI{5.2}{\micro\meter}$.
This turns out to match the typical \textit{mesh} size of $2-\SI{3}{\micro\meter}$ reported in \cite{Lieleg2009} quite well and is thus chosen as the default value for most of the simulations conducted in this work (once again refer to~\figref{fig::particle_mobility_setup_networks} for a visual comparison of model and real hydrogels).
The densest network to be considered in this work is given as~$n_\text{VP}=600$ and thus results in a high fiber volume fraction of~$\bar V_\text{f} \approx 1.8\%$ and cell diameters in the range of approx.~$1.0-\SI{2.5}{\micro\meter}$.
This second value of the network parameter~$n_\text{VP}=600$ is motivated by the example of the human amniotic basal membrane \cite{Yurchenco1987}, which appears to be much denser than the one considered above.
One example of the resulting model network is shown in~\figref{fig::particle_mobility_setup_networks_comparison_highfibervolfrac}.

In order to investigate and average out the influence of the specific network geometry, 5 different realizations of the random network generation process will be considered as input for the simulations for each -- otherwise identical -- set of input parameters in Sections~\ref{sec::particle_mobility_results_charge} and \ref{sec::particle_mobility_results_varying_stiffness}, respectively.
In contrast to the spherical particle, which will be discussed next, the fibers are assumed to be athermal because the lengths of the individual fiber segments are generally much smaller than their persistence length~$\ell_\text{p} \approx \SI{39}{\micro\meter}$ (resulting from the parameters for the bending stiffness given above and the thermal energy at room temperature~$k_\text{B}T \approx \SI{4e-3}{\atto\joule}$) and thermal undulations will therefore be negligible.
However, for another network architecture with longer fiber segments or a different species of thinner and thus more flexible fibers such as mucin or F-actin, this effect might become important and may be included in the novel computational model by means of the formulation proposed in \cite{Cyron2010}.

Finally, the spatial discretization of the fibers makes use of the geometrically exact Hermitian Simo-Reissner beam elements proposed in~\cite{Meier2017c}, which are mainly based on the well-established element formulation proposed by Crisfield and Jeleni\'c \cite{jelenic1999,crisfield1999}.
The applied centerline interpolation using cubic Hermite polynomials ensures both high accuracy in terms of spatial approximation and a $C^1$-continuous geometry representation.
This is particularly important for smooth contact kinematics and smooth interaction force distributions as has been shown in the context of beam-beam interactions in \cite{GrillSSIP,GrillPeelingPulloff}.
When creating the finite element discretization for each of the random network configurations, a default element length~$l_\text{ele} = \SI{1}{\micro\meter}$ is used for each straight fiber segment and one additional shorter beam element is created for the remainder of the random segment length if needed.
Based on the previous experience (from challenging scenarios with large deformations such as presented in \cite{GrillSSIP}) with this kind of beam finite element featuring fourth order spatial convergence, this is considered to be a sufficiently fine spatial resolution for the magnitude of deformations observed throughout this computational study, such that the influence of the spatial approximation error on the results is expected to be negligible.
To conclude, this strategy typically leads to a number of fiber segments~$n_\text{f} \approx 800$, a number of beam elements~$n_\text{ele} \approx 1600$, and a number of nodes~$n_\text{node} \approx 4000$ for~$n_\text{VP} = 60$.
Likewise, for the densest network with~$n_\text{VP} = 600$, we obtain $n_\text{f} \approx 6200$, $n_\text{ele} \approx 6200$, and $n_\text{node} \approx 18700$.

\subsection{Spherical particle}
The particle is modeled as a rigid sphere and is therefore uniquely described by its midpoint position~$\vr_\text{p}(t) \in \MR^3$ as a function of the time~$t$ as well as its diameter~$D_\text{p} = \SI{1}{\micro\meter}$ that is chosen to comply with the experimental studies presented in \cite{Lieleg2010,Arends2013}.
As in several of the aforementioned previous computational studies (e.g.,~\cite{Hansing2016,Hansing2018}), the Brownian dynamics of the particle is modeled by the Langevin equation (see e.g.,~\cite{Doi1988}), including the stochastic thermal as well as viscous drag forces that are repeated here for the reader's convenience.
The velocity-proportional drag force~$\vf_\text{s,visc}$ implicitly models a quiescent surrounding fluid and makes use of the friction coefficient of a sphere~$\gamma = 3 \pi \eta D_\text{p}$ according to Stokes:
\begin{equation}
  \vf_\text{s,visc} = \gamma \vI_{3\times 3} \, \dot \vr_\text{p}.
\end{equation}
Here,~$\dot \vr_\text{p}$ denotes the particle velocity, and the fluid viscosity is chosen as $\eta = \SI{1}{\milli\pascal \second}$, which corresponds to the viscosity of water at room temperature and has been found to be in good agreement with experimental tracer studies of freely diffusing particles \cite{Arends2013}.
Following the consistent modeling of the Brownian dynamics of slender biopolymers within the space- and time-discrete theoretical framework of the nonlinear finite element method according to \cite{Cyron2010}, the stochastic thermal forces acting on the sphere are given as:
\begin{equation}\label{eq::sphere_stochastic_force}
  \vf_\text{s,stoch} = \sqrt{2 k_\text{B} T} \,\gamma \vI_{3\times 3} \, \pdiff{\mathbf{\mathcal{W}}(x,t)}{{x}t}.
\end{equation}
Here, the thermal energy is set to~$k_\text{B} T = \SI{4.1e-3}{\atto\joule}$ corresponding to room temperature and the last term describes the space-time white noise resulting from a two-dimensional Wiener process~$\mathcal{W}(x,t)$.
Both contributions from viscous and stochastic forces on the spherical particle are added to the total virtual work of the system, which other than that, includes the contributions from internal elastic forces of the fibers and the contributions from the particle-fiber interactions that will be discussed next.
Afterward, the temporal discretization and time stepping scheme will be presented in \secref{sec::particle_mobility_model_setup_temporal_discretization}.

\subsection{Interactions between the particle and the fiber network}
\label{sec::particle_mobility_model_setup_interactions}
In view of the central research questions described above, the interactions between the spherical particle and the fiber network are the key component of this computational model.
In accordance with the sophisticated modeling of elastic fibers, these interactions are modeled at the level of individual particle-fiber pairs and are evaluated as a fully resolved, resulting line force distribution on the fiber.
At this point, it becomes clear that the sophisticated fiber model, including potentially large 3D deformations, which is a unique feature of this novel computational model, carries over to interaction modeling in the form of additional challenges to accurately and efficiently describe the sphere-fiber interactions for arbitrarily deformed fiber configurations and mutual orientations.
Since the conclusion of previous experimental as well as computational studies is that the combination of repulsive steric and (attractive) electrostatic effects is the main reason for the effective selective filtering of biological hydrogels (see e.g.,~\cite{Lieleg2010,Arends2013,Hansing2018}), both interaction types are accounted for in the present model.

To the best of the authors' knowledge, the problem of a beam interacting with a rigid sphere via both repulsive steric and (attractive) electrostatic effects has not been considered in the literature before.
The key idea of our modeling approaches is to consider the sphere-beam interaction as a special, simpler case of the beam-beam interaction.
Starting from the formulation for molecular interactions proposed in \cite{GrillSSIP} and the contact formulation proposed in \cite{meier2016}, the following procedure basically replaces one of the beams of an interaction pair by a rigid sphere.
The resulting formulations, derived for both contact and electrostatic interactions between a rigid sphere and a beam, are summarized below.
Note that hydrophobic interactions are not included here due to the lack of a proper fundamental understanding and modeling approach.
Assuming that this effect can be described by an effective interaction potential, it could however be incorporated in a very similar manner as the electrostatic interactions and would be an interesting future extension of the present model.

\subsubsection{Electrostatic interactions}
Generally, the two-body interaction potential of a beam-sphere pair~$\Pi_{ia}$ shall be described as:
\begin{align}\label{eq::pot_ia_beam_sphere}
  \Pi_{ia} = \int \limits_0^{l} \tilde{\pi}_\text{section-sphere}(\vr_\text{b-s},\vpsi_{\text{b-s}}) \dd s,
\end{align}
where a section-sphere interaction potential law~$\tilde{\pi}_\text{section-sphere}$ has been introduced.
Such a reduced interaction law is an analytical description of the effective interaction of one cross-section of the beam with the rigid sphere and will be further specified below.
In general, $\tilde{\pi}_\text{section-sphere}$ will depend on the relative distance vector~$\vr_\text{b-s}$ of the section midpoint~$\vr_\text{b}(s)$ and sphere midpoint~$\vr_\text{s}$ as well as the relative orientation expressed by the relative rotation vector~$\vpsi_{\text{b-s}}$.
In order to arrive at the total two-body interaction potential~$\Pi_{ia}$, the section-sphere interaction potential~$\tilde{\pi}_\text{section-sphere}$ is then numerically integrated along the arbitrarily deformed centerline curve of the beam.
Here, $s \in [0,l]$ denotes the arc-length parameter, which is defined in the stress-free, initial configuration of the beam's centerline curve.
This general approach is in close analogy to the concept of section-section interaction potential (SSIP) laws introduced in \cite{GrillSSIP}.

Since it is the most relevant specific example for this study, we now consider the case of Coulomb interaction.
Again in close analogy to the careful choice of the SSIP law for this kind of interaction in \cite{GrillSSIP}, the following section-sphere interaction potential law~$\tilde{\pi}_\text{section-sphere}$ is obtained:
\begin{align}
  \tilde{\pi}_\text{section-sphere} = \lambda_\text{b} \, Q_\text{s} \, \Phi(d) \quad \text{with~~} d = \norm{ \vr_\text{b} - \vr_\text{s}}.
\end{align}
Here, $\lambda_\text{b}$ denotes the line charge density of the beam, $Q_\text{s}$ denotes the total charge of the sphere, and $\Phi(d)=C_\text{elstat} \, d^{-1}$ is the well-known Coulomb potential with its inverse distance dependency and constant prefactor~$C_\text{elstat}$.
Following the theoretical considerations in \cite{GrillSSIP}, the section-sphere interaction potential law~$\tilde{\pi}_\text{section-sphere}$ is expressed solely by the scalar separation of the section and sphere midpoint positions.
According to the detailed study of the accuracy of this simple resulting interaction law in \cite{GrillSSIP}, the intentional neglect of the orientation dependence~$\vpsi_{\text{b-s}}$ turned out to be a reasonable approximation in the case of circular, homogeneous cross-sections and long-range interactions as considered here.

The required variation of Equation~\eqref{eq::pot_ia_beam_sphere} for obtaining the corresponding virtual work contribution, the subsequent spatial discretization of the beam centerline for arriving at the discrete element force vector for both the beam and the sphere as well as the consistent linearization of these terms will be omitted here for the sake of brevity.
They follow directly from substituting the respective expressions in the equations for the SSIP approach as presented in \cite{GrillSSIP}.

Note that arguably the most critical limitation of this computational model is the use of a Coulomb interaction potential law, which neglects the presence of counterions in the electrolyte solution and the associated screening of charges that in turn significantly reduces the range of electrostatic interactions in biological hydrogels.
This can be seen as a pragmatic, simplified model chosen due to the current lack of a sphere-beam or beam-beam interaction formulation for screened electrostatic interactions, and the implications will be discussed in the following paragraph.\\

\noindent\textit{Discussion of neglecting screening effects.}\\
Regarding the impact of this model assumption on the results of this computational study, it is expected that the effect of charges in general and the range of interaction in particular is overestimated and that applying the simple Coulomb interaction model is thus inadequate for predicting the particle mobility in a quantitatively correct manner.
However, the qualitative behavior of the particle and the trends in the statistical quantities of interest such as the mean squared displacement (MSD) of the particle over time for varying charge density is expected to be meaningful and thus allow for both valuable insights in the biophysical system behavior and mechanisms as well as a first proof of principle for this novel computational modeling approach in general.
Based on the experimental observations and in anticipation of the obtained simulation results, this reasoning is supported by the fact that the most effective trapping mechanism is the one of a persistent, strongly adhesive contact between the particle and the oppositely charged network fibers, which is a scenario with minimal surface separation and thus minimal screening effect.
Therefore, the behavior of a particle with medium to large distance to the nearest fibers is thus expected not to be reproduced correctly by the Coulomb interaction model, whereas in contrast, the practically much more important regime of small separations should be represented with sufficient accuracy in order to allow for the aforementioned analysis of trends and basic mechanisms.
In order to still account for screening charges, the cutoff distance of the interaction is set to~$r_\text{cutoff}=\SI{2}{\micro\meter}$, which is defined via the separation of the sphere and fiber midpoint position and thus effectively neglects any interaction forces beyond particle-fiber surface separations of~$g_\text{cutoff} \approx \SI{0.96}{\micro\meter}$.
Finally, despite the fact that the development of a screened electrostatic interaction formulation for instance based on the Debye-H\"uckel approximation of the Poisson-Boltzmann theory would go beyond the scope of this computational study, it is clearly considered an important and promising future extension of the present model that should be used for both subsequent verification of the drawn conclusions and specific analysis of the influence of salt concentration as well as ion-specific effects observed in experiments \cite{Arends2013}.\\

The parametrization of the electrostatic interaction model used throughout this study is given as follows.
Based on assuming the dielectric permittivity of water (at room temperature) for the surrounding fluid, the constant prefactor of the Coulomb interaction potential law is obtained as~$C_\text{elstat} \approx \SI{1.12e2}{\atto\joule \micro\meter \per\square\femto\coulomb}$.
As a first step and in accordance with all previous computational studies, the surface charge density of the fibers as well as that of the particle is assumed to be homogeneous and constant along the fibers.
In view of the complex, inhomogeneous molecular architecture and thus charge distribution of individual biopolymer filaments and moreover the complex constitution of a real biological hydrogel, this model assumption is once again expected to have a significant influence on the quantitative accuracy of the results.
However, it should still allow for qualitative analysis of trends and mechanisms, as argued above in the context of the electrostatic interaction model.
A potential improvement on this point is rather a question of detailed experimental fiber characterization and model parametrization than method development, because the sphere-beam interaction model proposed above is capable of describing varying line charge distributions along the filaments.
As a first step, however, a constant, homogeneous line charge density of the fibers~$\lambda_\text{f} = \SI{-0.25}{\femto\coulomb \per \micro\meter}$ is assumed throughout this article and the positive surface charge of the particle will be varied in \secref{sec::particle_mobility_results_charge} to study its influence on the particle mobility.
Finally, a total of 10 integration points per beam element are used to evaluate the contributions of the electrostatic line force distribution along the fibers by means of Gaussian quadrature.

\subsubsection{Repulsive steric interactions}
The line-to-line contact formulation proposed in \cite{meier2016} effectively precludes any noticeable penetration of fibers for arbitrary mutual orientations and deformations.
It thus serves as the starting point for the dimensionally reduced case of beam-to-rigid sphere contact, which is outlined as follows:
Postulating a beam-sphere penalty force law as a linear function of the minimal surface-surface separation (i.e.~gap)~$g_\text{b-s}$ with constant scalar prefactor (i.e.~penalty parameter)~$\varepsilon_\text{b-s}$ yields the two-body interaction potential:
\begin{align}\label{eq::pot_ia_beam_sphere_penalty_contact}
    \Pi_{\text{c$\varepsilon$,b-s}} = \frac{1}{2} \varepsilon_\text{b-s} \int \limits_0^{l} \langle g_\text{b-s}(s) \rangle^2 \dd s.
\end{align}
Here, the crucial difference to the beam-beam scenario lies in the way the gap~$g$ is computed.
Whereas a unilateral closest-point projection is required in the beam-beam scenario, the problem simplifies significantly in the case of a rigid sphere, because the minimal surface separation between the current beam cross-section and the sphere may be expressed in good approximation as:
\begin{align}
  g_\text{b-s}(s) = \norm{ \vr_\text{b} - \vr_\text{s} } - R_\text{b} - R_\text{s}.
\end{align}

As in the previous section, the required variation of Equation~\eqref{eq::pot_ia_beam_sphere_penalty_contact} for obtaining the corresponding virtual work contribution and subsequent spatial discretization of the beam centerline for arriving at the discrete element force vector for both the beam and the sphere as well as the consistent linearization of these terms follows in close analogy to the beam-beam scenario and will therefore not be presented here.

The parametrization of this contact model is chosen as follows:
Throughout this study, a constant penalty parameter~$\varepsilon_\text{b-s} = \SI{100}{\pico\newton \micro\meter^{-1}}$ is chosen, which turned out to be sufficiently large such that the maximum penetration of particle and fiber is smaller than $5\%$ of the fiber diameter~$D_\text{f}$ even in the most challenging scenarios of strongly adhesive contact and sudden stochastic forces on the particle in the direction toward the fiber.
Finally, 15 collocation points per beam element are used for the evaluation of the contact line force distribution along the fibers.\\

\noindent\textit{Remark on fiber-fiber interactions.}\\
At the end of this section, note that contact and electrostatic interactions between fibers are not considered here because it turns out that their mutual separations and orientations are almost entirely determined by the rigid connections of their endpoints at the network vertices.
As mentioned earlier, beam-beam interaction formulations for both steric repulsive and electrostatic interactions are readily available and could be directly added as a future extension of this model, however, at the cost of an increased computational effort.

\subsection{Temporal discretization}
\label{sec::particle_mobility_model_setup_temporal_discretization}
In addition to the already discussed spatial discretization of the fibers via beam finite elements, the problem is discretized in time via an implicit direct time stepping scheme.
Specifically, the resulting system of first-order stochastic%
\footnote{Recall that our Brownian dynamics model includes the stochastic thermal forces acting on the spherical particle as stated in Equation~\ref{eq::sphere_stochastic_force}.}
partial differential equations in time is discretized by means of a Backward Euler scheme, as proposed in \cite{Cyron2010}.
Starting from a default time step size of~$\Delta t = 10^{-3} \SI{}{\second}$, an adaptive time stepping scheme is applied, which is especially important for resolving the highly nonlinear dynamics during (adhesive) contact interactions and potentially large sudden changes in the magnitude and direction of the thermal forces.
The total simulation time per run is set to~$t_\text{end} = \SI{20}{\second}$, which generally leads to a required number of time steps in the range of~$n_\text{step} \in \, [ \SI{2e5}{}, \SI{3e5}{} ]$.
In order to account for the stochastic nature of the thermal forces driving the particle motion, generally two or more random realizations for each -- otherwise identical -- set of input parameters (i.e.,~identical, random network geometry and interaction parameters) are computed and considered in the analyses presented in \secref{sec::particle_mobility_results}.

\subsection{Boundary conditions}
\label{sec::particle_mobility_model_setup_boundary_conditions}
As already mentioned in the context of the network modeling (\secref{sec::particle_mobility_model_setup_fiber_network}), the concept of a representative volume element is used to limit the influence of artificial boundary effects.
For this purpose, periodic boundary conditions are applied at each side of the simulation box shown in~\figref{fig::particle_mobility_network_600VP}.
Thus, once the particle as well as (parts of) the fibers leave the representative volume element at any side, they reenter it on the opposite side.
Moreover, the steric as well as electrostatic interactions are also evaluated across periodic boundaries.
In the majority of the simulations considered in \secref{sec::particle_mobility_results}, no other boundary conditions are applied.
However, in a certain batch of simulations, the entire network of fibers will be fixed by means of Dirichlet boundary conditions in order to serve as a reference solution mimicking the limiting case of rigid fibers.

\subsection{Postprocessing and quantities of interest}
Since this study focuses on the (hindered) diffusive mobility of particles, the most important raw data obtained from the simulations is the particle midpoint position in every time step.
From this point on, the postprocessing procedure is equivalent to experiments that track the motion of individual fluorescent tracer particles (e.g.,~\cite{Lieleg2009,Lieleg2010,Arends2013}).
Based on the discrete time sequence of particle positions, the mean squared displacement (MSD) $<\Delta r_\text{p}^2(\tau)>$ is computed for any desired time interval (that can be observed in a given simulation run)~$\tau \in \, [\Delta t, t_\text{end} ]$ as follows:
\begin{equation}
  <\Delta r_\text{p}^2(\tau)> \; = \frac{1}{N_{\tau}} \sum_{i=0}^{N_{\tau}} \left( \vr_\text{p}(i \Delta t + \tau) - \vr_\text{p}(i \Delta t) \right)^2.
\end{equation}
Here, $N_{\tau}$ denotes the number of all distinct (but possibly overlapping) time intervals~$\tau$ obtained for one simulation run.
Given that the number of \textit{independent} samples obtained for large time intervals is naturally limited, only the first 10\% of the maximal possible time intervals, i.e.,~$\tau \in \, [\Delta t, 0.1 \cdot t_\text{end} ]$, will be considered in the statistical analyses.
However, the remaining data is included and indicated by a gray background in all the MSD plots to be presented.
Moreover, the mean and standard deviation of the MSD obtained for several realizations of the random network geometry as well as several realizations of the stochastic process of thermal forces will be considered.
As the majority of the considered scenarios shows a subdiffusive behavior, the MSD curves over the time interval will be presented and discussed instead of the (apparent) diffusion constant, which obviously depends on the considered time interval and could still be computed from the MSD curves if desired.
Other simulation results, such as the resulting axial strains of the fibers, will be presented and discussed for a few specific investigations wherever needed for interpreting the system behavior.

\section{Results and discussion}
\label{sec::particle_mobility_results}
The following discussion of simulation results is divided into three parts.
First, in \secref{sec::particle_mobility_results_steric_hindrance_only}, we study the influence of collisions between the particle and the fiber network on the particle mobility.
Second, the effect of additional attractive electrostatic interactions between the particle and the fiber network will be investigated in \secref{sec::particle_mobility_results_charge}.
Lastly, \secref{sec::particle_mobility_results_varying_stiffness} analyzes the special role of fiber stiffness in the case of the most effective hindrance mechanism observed in the simulations.

\subsection{The effect of solely repulsive steric interactions}
\label{sec::particle_mobility_results_steric_hindrance_only}
\begin{figure}[htpb]%
  \centering
  \subfigure[]{
    \includegraphics[width=0.53\textwidth]{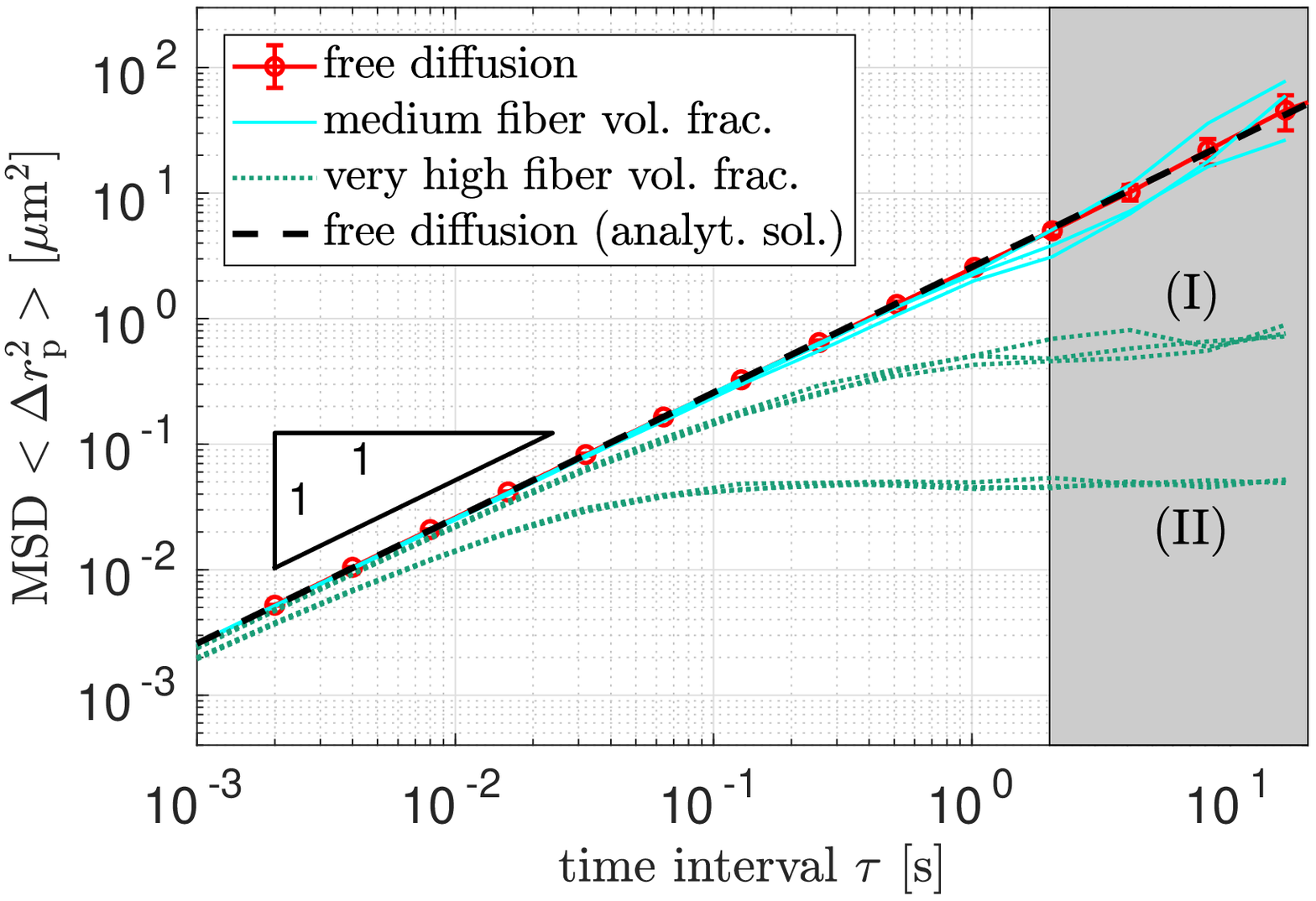}
    \label{fig::particle_mobility_msd_over_timeinterval_steric_hindrance_only}
  }
  \subfigure[]{
    \includegraphics[width=0.43\textwidth]{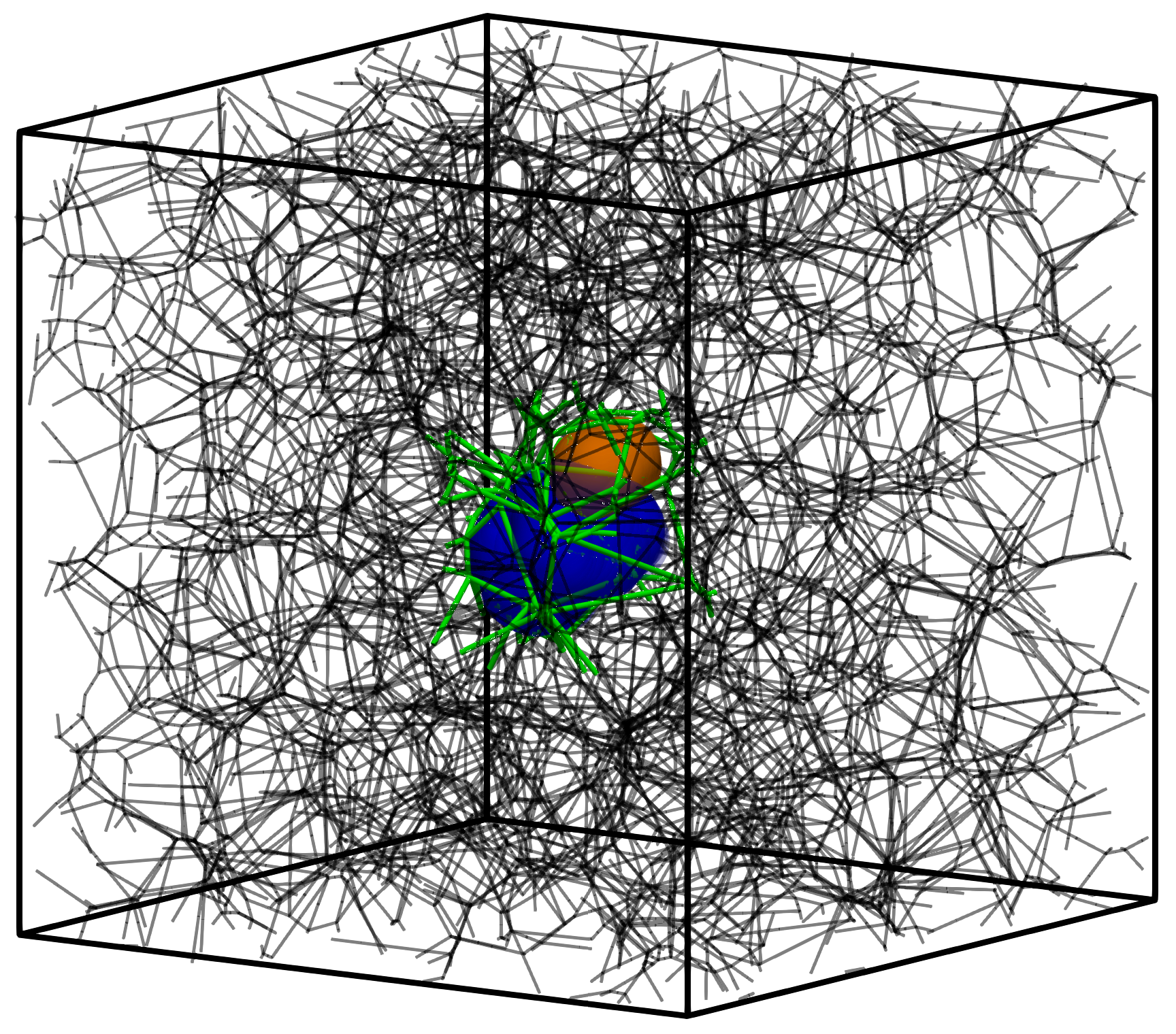}
    \label{fig::particle_mobility_VP600-trap-compartments-2}
  }
  \subfigure[]{
    \includegraphics[width=0.35\textwidth]{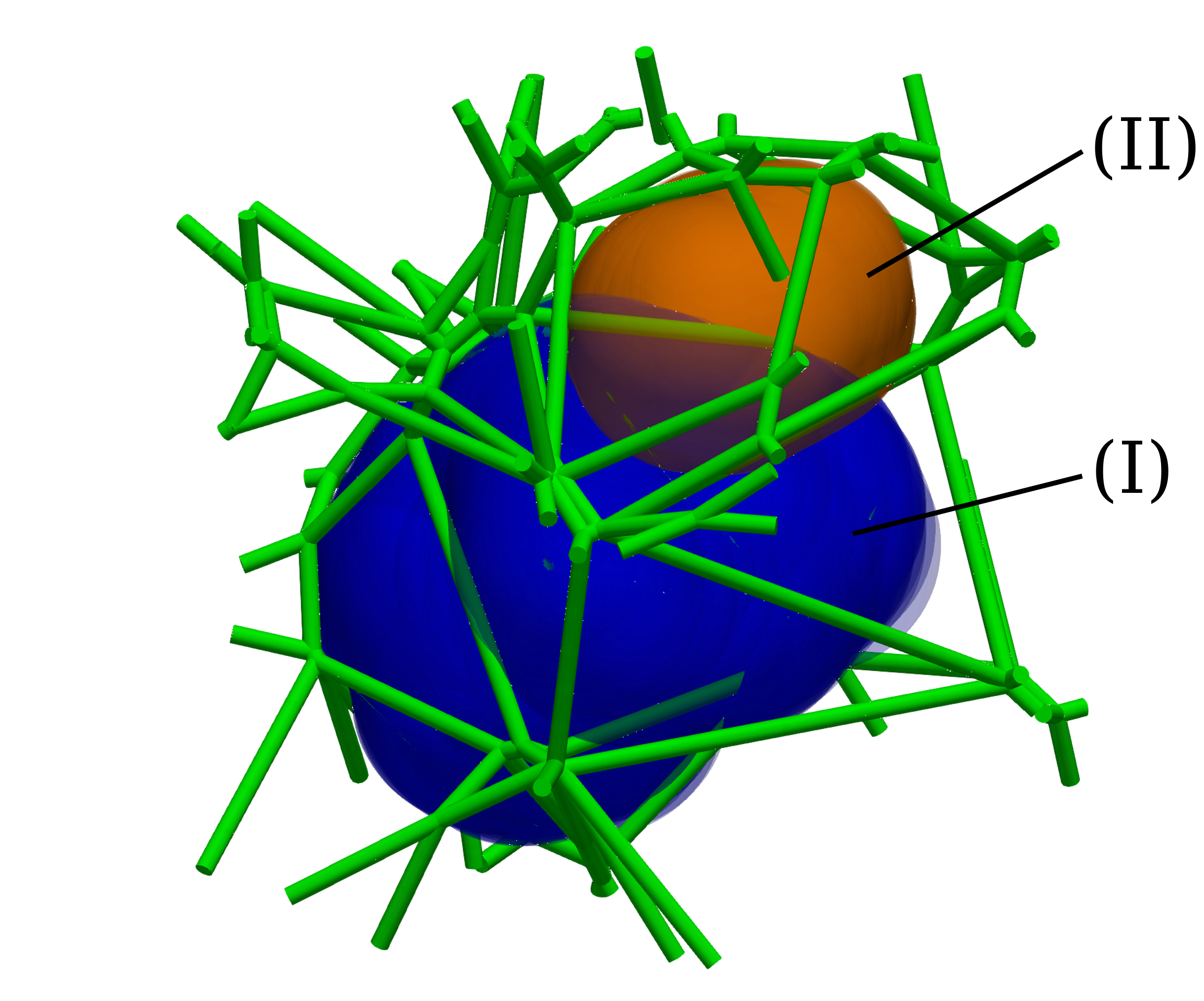}
    \label{fig::particle_mobility_VP600-trap-compartments-2-detail}
  }
  \caption{Analysis of the particle mobility in presence of solely repulsive steric interactions with the fiber network.
  (a) Mean squared displacement (MSD) of the particle $<\Delta r_\text{p}^2>$ as a function of the time interval~$\tau$: Mean and standard deviation over ten random realizations for the case of free diffusion (red line with circles and error bars) and three individual random realizations for a medium fiber volume fraction~$\bar V_\text{f} \approx 0.4\%$ (cyan lines).
  The analytical solution for the case of free diffusion is plotted as a reference (black dashed line).
  In addition, two sets (I) and (II) of three individual random realizations each for a very high fiber volume fraction~$\bar V_\text{f} \approx 1.8\%$ (green dotted lines) are included.
  To demonstrate the extreme effect of caged particles, all fibers have been fixed in space for all realizations with~$\bar V_\text{f} \approx 1.8\%$.
  The gray background indicates the range of time intervals above 10\% of the simulation time, where only few independent samples are available for computing the MSD.
  (b) Network with very high fiber volume fraction~$\bar V_\text{f} \approx 1.8\%$ and overlay of all observed particle positions in one simulation run corresponding to either the high MSD plateau value (I, blue) or the low value (II, orange).
  (c) Magnified detail showing the two compartments with irregular polygonal shape that the particle cannot leave.}
  \label{fig::particle_mobility_results_steric_hindrance_only}
\end{figure}

To begin with, the double-logarithmic plot of the particle's  MSD~$<\Delta r_\text{p}^2>$ as a function of the time interval~$\tau$ shown in \figref{fig::particle_mobility_msd_over_timeinterval_steric_hindrance_only} confirms the validity of the applied Brownian dynamics model, because the results obtained for free diffusion of the particle (red line with circles and error bars) excellently match the analytical reference solution (black dashed line).
Moreover, the standard deviation of the mean over ten random realizations as indicated by the error bars is negligible within the first 10\% of the time interval range (indicated by the white background) that is considered in the following analyses.\\

\noindent\textbf{Steric hindrance is insignificant as long as particle diameters are smaller than the smallest mesh sizes.}\\
Turning to the influence of repulsive steric interactions, i.e., collisions between the particle and the fiber network, \figref{fig::particle_mobility_msd_over_timeinterval_steric_hindrance_only} shows that a medium fiber volume fraction~$\bar V_\text{f} \approx 0.4\%$ (cyan lines, see \figref{fig::particle_mobility_network_60VP} for an example of the network) has almost no perceptible influence on the MSD.
Only above approximately $\tau=\SI{0.5}{\second}$ is a very subtle subdiffusive behavior observable for the three individual realizations plotted here, which indicates the occurrence of a few collisions if the particle travels over longer periods of time.
On the one hand, this behavior of almost free diffusion of the particle is expected from the range of cell diameters $2.7 - \SI{5.2}{\micro\meter}$ of this irregular polygonal network compared to the particle diameter~$D_\text{p} = \SI{1}{\micro\meter}$.
Moreover, previous experimental studies have made very similar observations for almost neutral particles or high salt concentrations that effectively shield any electrostatic interactions~\cite{Lieleg2009,Lieleg2010,Arends2013}.
However, this is the first computational study with a realistic, irregular polygonal fiber network geometry and thus an important confirmation that the effect of solely steric hindrance is indeed negligible in this regime, where the ratio of network mesh size(s) and particle diameter is greater than one.\\

\noindent\textbf{Particles with diameters in the range of the mesh sizes are caged in polygonal compartments of random size and show confined diffusive behavior.}\\
To push this to the limit where purely steric hindrance and thus filtering by particle size will become significant, a ten times larger number of Voronoi points~$n_\text{VP}=600$ corresponding to a fiber volume fraction of~$\bar V_\text{f} \approx 1.8\%$ and cell diameters of $1.0 - \SI{2.5}{\micro\meter}$, i.e., in the order of the particle diameter~$D_\text{p}=\SI{1}{\micro\meter}$, have been applied.
An example of the resulting network architecture is illustrated in \figref{fig::particle_mobility_network_600VP}.
To probe the most extreme effect of steric hindrance, the fibers are completely fixed in space for these two sets of three random realizations each.
The results are plotted in \figref{fig::particle_mobility_msd_over_timeinterval_steric_hindrance_only} (green dotted lines).
One of the sets (I) shows close to normal diffusive behavior on very small time scales up to approximately~$\tau=\SI{0.05}{\second}$ and eventually reaches a plateau value of~$<\Delta r_\text{p}^2> \approx \SI{0.7}{\micro\meter^2}$ beyond~$\tau \approx \SI{1}{\second}$.
The other set (II), which are random realizations using the identical fiber network geometry, but a different initial position of the particle, shows a significantly subdiffusive behavior already for the smallest considered time intervals and quickly reaches a plateau value of~$<\Delta r_\text{p}^2> \approx \SI{0.05}{\micro\meter^2}$ for any time interval longer than~$\tau \approx \SI{0.1}{\second}$.

This is an expected behavior for the diffusion of particles in an irregular network with cell diameters of the same order as the particle diameter and therefore randomly connected sufficiently large cells that together form a polygonal volume surrounding the initial position of the particle, which the particle cannot leave under any circumstances.
The specific compartments that the particle is able to explore starting from either of the two initial positions in the identical network are illustrated in \figref{fig::particle_mobility_VP600-trap-compartments-2} and \ref{fig::particle_mobility_VP600-trap-compartments-2-detail}.
Here, the overlay of all particle positions throughout the entire simulation is shown in dark blue for the case of the higher MSD plateau value (I) and in orange for the case of the lower MSD plateau value (II) observed in \figref{fig::particle_mobility_msd_over_timeinterval_steric_hindrance_only}.
This behavior is known as confined diffusion and has been theoretically described and experimentally observed e.g.,~in the context of studying cadherin molecule mobility in plasma membranes \cite{Kusumi1993}.
As discussed for instance in \cite{Witten2017}, such a filtering mechanism based on size clearly has a very effective selectivity and is applied by organisms to strictly preclude the access of any objects larger than the characteristic mesh size, e.g.,~of the nasal mucous membrane.\\

However, there are still open questions concerning the mobility of (medium to) large objects in biological hydrogels taking into account the continuous reorganization of biopolymer networks based on both (de-)polymerization and the transient nature of crosslinks.
Such a transport mechanism for relatively large particles has been observed in experiments \cite{Lai2007} and recently been investigated also in a theoretical and computational model \cite{Goodrich2018}, where crosslink binding dynamics are influenced by the diffusing particle.
It is also suggested that this kind of mechanism could play a role in the selective permeability of the nuclear pore complex, for which the governing principles are still under debate (see e.g.,~the review article \cite{Witten2017}).
Replacing the Voronoi-type network in the present computational model by that of a transiently cross-linked, self-assembled network driven by Brownian motion \cite{Cyron2013phasediagram,Cyron2013a} is thus considered a promising future step.

\subsection{The effect of additional attractive electrostatic interactions}
\label{sec::particle_mobility_results_charge}
In addition to repulsive steric interactions, attractive electrostatic interactions due to uniformly distributed, opposite charges on the particle and the fibers will now be considered.
This has been confirmed to be the most effective hindrance mechanism for particles smaller than the mesh sizes both in experiments (e.g.,~\cite{Lieleg2009,Lieleg2010,Arends2013}) and simulations (e.g.,~\cite{Hansing2016,Hansing2018}).
Throughout this section, we thus keep the fiber volume fraction fixed at the medium value~$\bar V_\text{f} \approx 0.4\%$, which has been shown to not have a noticeable influence on the particle's MSD in \figref{fig::particle_mobility_msd_over_timeinterval_steric_hindrance_only} (cyan lines).\\

\begin{figure}[htb]%
  \centering
  \subfigure[]{
    \includegraphics[width=0.48\textwidth]{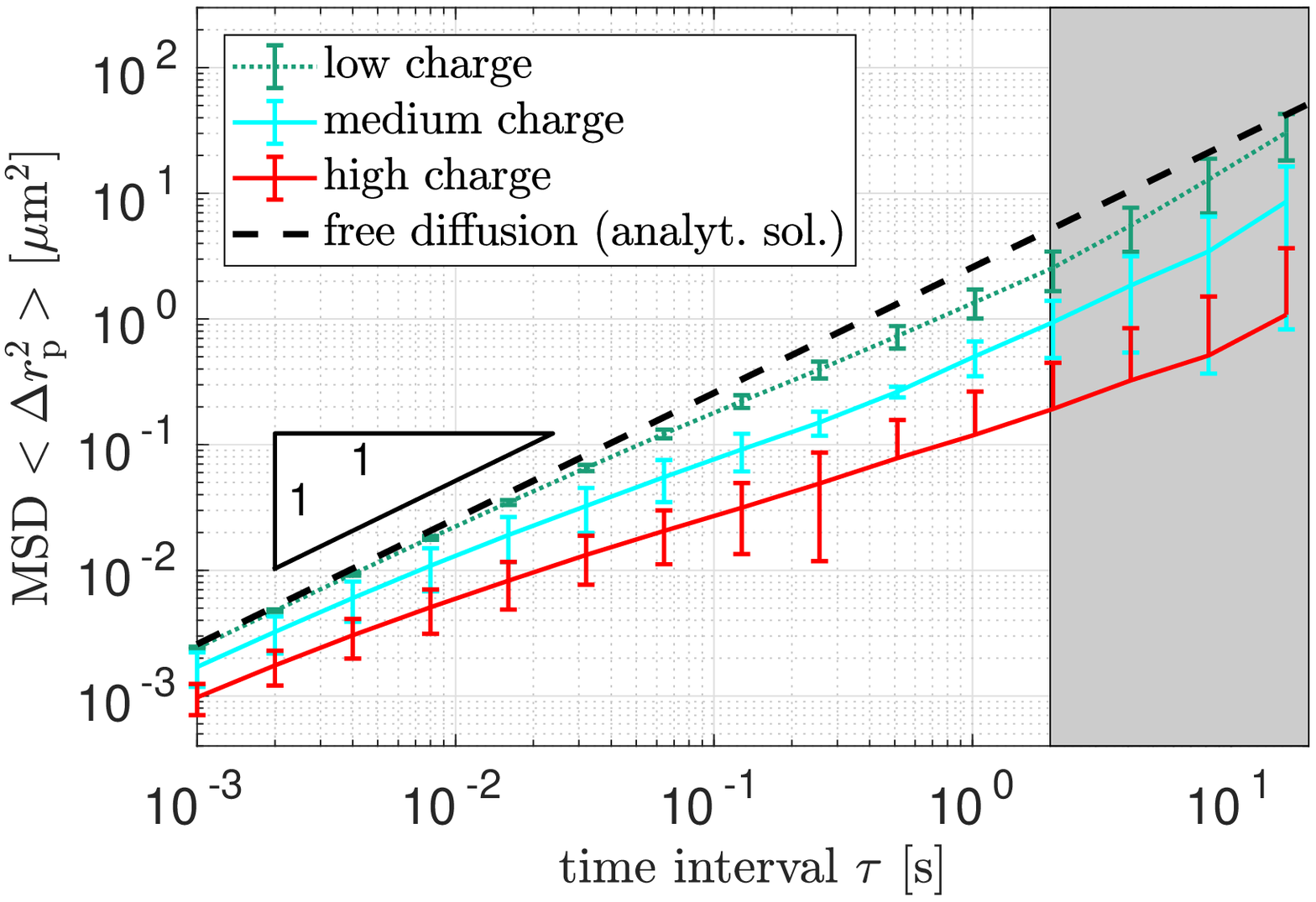}
    \label{fig::particle_mobility_msd_over_timeinterval_charge_mean_stddev}
  }
  \subfigure[]{
    \includegraphics[width=0.48\textwidth]{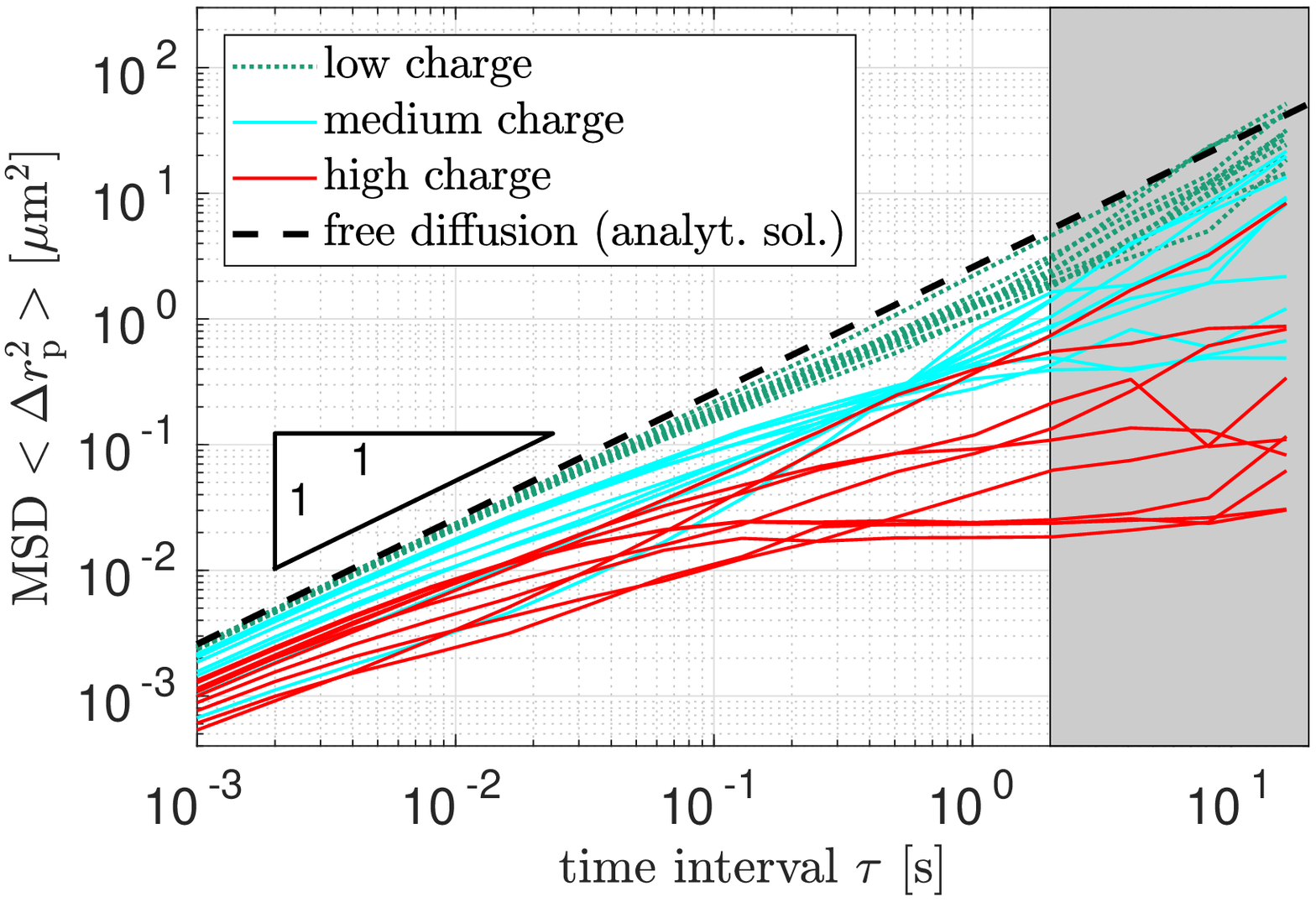}
    \label{fig::particle_mobility_msd_over_timeinterval_charge_allrealizations}
  }
  \caption{Analysis of the hindrance of particle mobility due to attractive electrostatic interactions with the fiber network (in addition to repulsive steric interactions).
  Mean squared displacement (MSD) of the particle $<\Delta r_\text{p}^2>$ as a function of the time interval~$\tau$:
  (a) Mean and standard deviation over five random network geometries and two random realizations each for three different values of the particle's surface charge:
  Low charge~$Q_\text{p} = \SI{0.125e-3}{\femto\coulomb}$ (green), medium charge~$Q_\text{p} = 10^{-3} \, \SI{}{\femto\coulomb}$ (cyan), and high charge~$Q_\text{p} = \SI{8e-3}{\femto\coulomb}$ (red).
  (b) All corresponding individual realizations.
  The medium fiber volume fraction~$\bar V_\text{f} \approx 0.4\%$ is identical for all of these realizations.
  The analytical solution for the case of free diffusion is plotted as a reference (black dashed line).
  The gray background indicates the range of time intervals above 10\% of the simulation time, where only few independent samples are available for computing the MSD.
  The bottom of the error bar is hidden for clarity wherever the corresponding value is negative.}
  \label{fig::particle_mobility_msd_over_timeinterval_charge}
\end{figure}

\noindent\textbf{On average, the degree of subdiffusion increases with the strength of attraction.}\\
The resulting MSD curves for a low (green), medium (cyan), and high (red) value of the particle's surface charge~$Q_\text{p}$ are shown in~\figref{fig::particle_mobility_msd_over_timeinterval_charge} and compared to the analytical reference solution for free diffusion (black dashed line).
As mentioned earlier, the fiber volume fraction~$\bar V_\text{f} \approx 0.4\%$ is identical for all simulation runs.
In particular, 5 different random network geometries with 2 random realizations of the stochastic forces each have been simulated for each of the three different charge values.
\figref{fig::particle_mobility_msd_over_timeinterval_charge_allrealizations} shows these 10 independent realizations for each particle charge value and the corresponding mean values and standard deviations are plotted in \figref{fig::particle_mobility_msd_over_timeinterval_charge_mean_stddev}.
On average, the degree of subdiffusion increases with the strength of attractive interaction, which has been suggested by previous experiments (e.g.,~\cite{Lieleg2009,Arends2013}), and has been confirmed by previous computational studies using ordered (e.g.,~\cite{Miyata2012,Hansing2016}) and unordered \cite{Hansing2018} arrays of straight, rigid, mutually orthogonal fibers.
Here, the smallest charge value~$Q_\text{p} = \SI{0.125e-3}{\femto\coulomb}$ leads to a very small degree of subdiffusion, which is in fact quite similar to the one observed in the limit of no charge shown in \figref{fig::particle_mobility_msd_over_timeinterval_steric_hindrance_only}.
In contrast, medium (cyan) and high charge values (red), which are a factor of 8 and 64 higher than the smallest charge value, significantly hinder the diffusion of the particles already on (very) small time scales~$\tau < \SI{0.1}{\second}$.
In this regime, the slopes of the MSD curves are however close to one, which suggests normal diffusive behavior with a decreased diffusion constant.\\

\begin{figure}[htpb]%
  \centering
  \subfigure[]{
    \includegraphics[width=\textwidth]{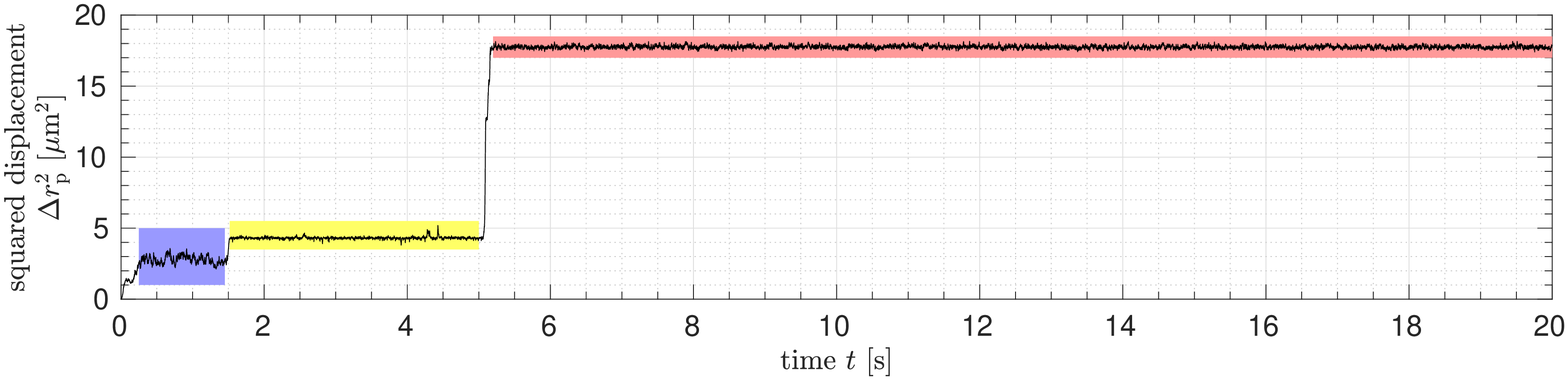}
    \label{fig::particle_mobility_sd_over_time_Q8_s3_1}
  }
  \subfigure[]{
    \includegraphics[width=0.5\textwidth]{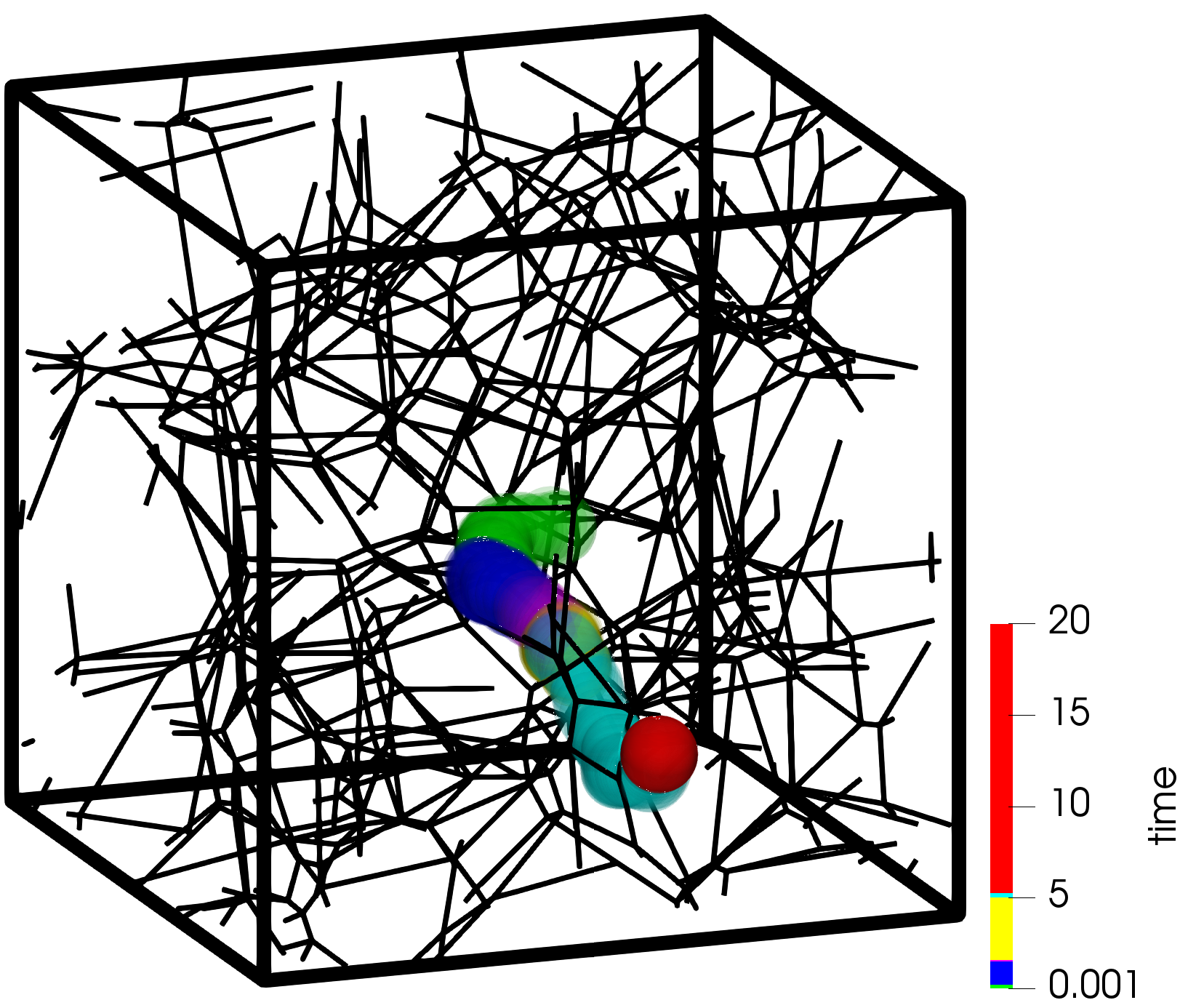}
    \label{fig::particle_mobility_Q8_s3_1-escape-events-jumps}
  }
  \subfigure[]{
    \includegraphics[width=0.4\textwidth]{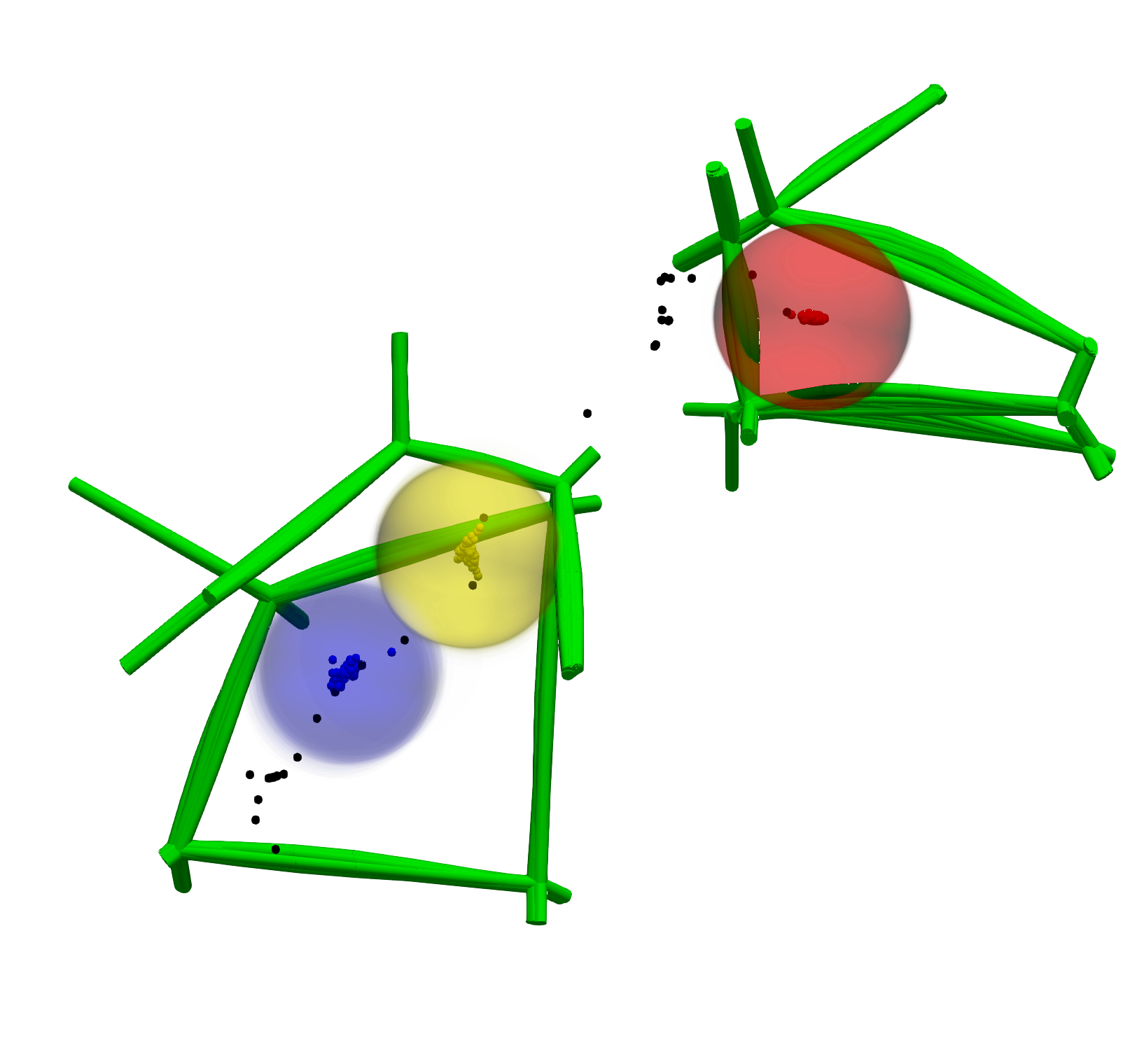}
    \label{fig::particle_mobility_Q8_s3_1-fiber-aggregate}
  }
  \caption{Simulation results for one realization with high charge~$Q_\text{p} = \SI{8e-3}{\femto\coulomb}$ and medium fiber volume fraction~$\bar V_\text{f} \approx 0.4\%$ that exhibits three distinct trapped states with sudden transitions (``jumps'') between them.
  (a) Squared displacement of the particle $\Delta r_\text{p}^2$ over the simulation time~$t$.
  (b) Particle trajectory with color (green, blue, pink, yellow, cyan and red) indicating the characteristic intervals of simulation time where the particle is either trapped (blue, yellow and red) or jumps between the trapped states (pink and cyan).
  (c) Detail view showing an overlay of the particle positions in the three distinct trapped states (blue, yellow and red) and an overlay of the (oppositely charged) fiber aggregates that are responsible for the effective trapping.
  The three colors (blue, yellow, and red) again match the corresponding time intervals in (a) and (b).}
  \label{fig::particle_mobility_jumps_three_trapped_states_Q8_s3_1}
\end{figure}

\noindent\textbf{The variability of MSD values and slopes increases for longer time intervals, which indicates that particles randomly switch between distinct motion patterns.}\\
This almost normal diffusive behavior with slope values close to one on very small time scales changes drastically for longer time intervals~$\tau > \SI{0.1}{\second}$, where the individual realizations exhibit slopes in the broad range from zero to one, and even some examples for superdiffusive behavior with a slope greater than one can be observed for medium charge values.
Such a significant increase in the variation of MSD values as well as slopes is in excellent agreement with the experimental results from \cite{Arends2013}.
Their subsequent analysis of squared displacement values over time revealed the existence of sudden trapping and escape events of a particle switching between an almost free, a loosely bound, and a tightly bound state.
This hypothesis is confirmed by the simulation results shown in \figref{fig::particle_mobility_sd_over_time_Q8_s3_1}, where very similar motion patterns can be observed.\\

\noindent\textbf{(Strongly) charged particles jump between local aggregates of fibers, i.e., opposite charges.}\\
While a causal link of trapped states and local aggregation of fibers, i.e., charge patches of opposite sign, has already been suggested in several experimental studies (e.g.,~\cite{Lieleg2009,Arends2013}), the limited spatio-temporal resolution of single particle tracking and imaging has so far precluded a direct proof.
The particle trajectory shown in \figref{fig::particle_mobility_Q8_s3_1-escape-events-jumps} and the detail view of the fiber aggregates corresponding to the three observed trapped states illustrated in \figref{fig::particle_mobility_Q8_s3_1-fiber-aggregate} clearly confirm this causality.
Related observations have been made in the recently published computational study \cite{Hansing2018}, which for the first time considered disorder in the still rigid, straight, mutually orthogonal fibers and described a resulting, so-called dense-region trapping.
In the present study, the random, irregular polygonal network structure, very similar to the one of real biological hydrogels (cf.~Figures~\ref{fig::particle_mobility_setup_networks} and \ref{fig::particle_mobility_setup_networks_comparison_highfibervolfrac}), allows for a more detailed analysis of this trapping mechanism.
Particularly, these fiber aggregates are found at network vertices and small polygonal faces in the network that have a lateral spatial extension smaller than the particle diameter.
\figref{fig::particle_mobility_jumps_three_trapped_states_Q8_s3_1} also suggests that the magnitude of the remaining thermal fluctuations observed in the squared displacement of the particle is a measure for the strength of the trapping and that this strength is proportional to the local fiber, i.e., charge density.
Finally, the results suggest that the stronger the particle is immobilized, i.e., the smaller the fluctuations in the squared displacement are, the longer the particle will remain at this location.\\

\noindent\textbf{The irregularity in the fiber/charge distribution gives rise to three distinct motion patterns, one of which allows the particle to travel via successive jumps.}\\
From a mechanical point of view, the spatially varying distribution of fibers and thus charges gives rise to a random 3D potential field to be explored by the particle (similar to the one used in our previous contribution \cite{Slepukhin2019} to study the effect of filament prestress in biopolymer networks).
The location and values of the (local) minima in the potential field as well as the potential barrier and paths connecting them strongly depend on the characteristic geometrical properties of the network such as connectivity, distribution of fiber segment lengths and the resulting distribution of cell and mesh sizes.
Considering real biological hydrogels further extends the list of crucial influencing factors to the type and fractions of load-carrying components and their specific surface charge distributions.
All of these factors that shape the characteristic potential landscape will determine whether and how the stochastic thermal excitation causes the particle to either (1) remain at one location being completely immobilized, (2) cycle between neighboring minima being restricted to a certain region, or (3) travel through the hydrogel via jumps -- potentially also over long distances.

\begin{figure}[htpb]%
  \centering
  \includegraphics[width=0.4\textwidth]{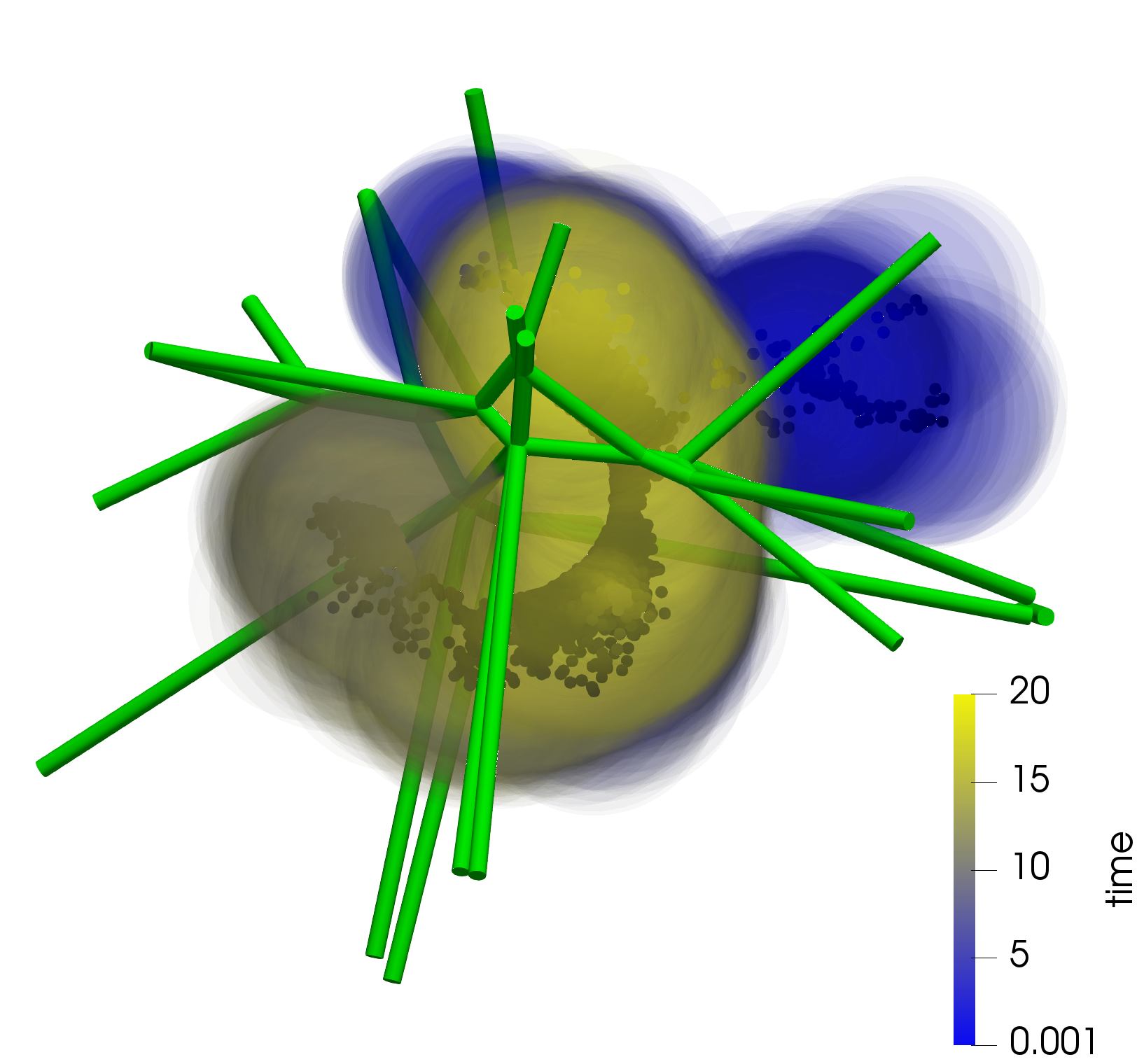}
  \caption{Particle trajectory for one realization with medium charge~$Q_\text{p} = \SI{1e-3}{\femto\coulomb}$ and medium fiber volume fraction~$\bar V_\text{f} \approx 0.4\%$ that smoothly cycles around a region of fiber/charge agglomeration. The continuous color scheme indicates the course of the simulation time~$t$.}
  \label{fig::particle_mobility_cycling}
\end{figure}

All three motion patterns (1)--(3) can be identified among the realizations shown in~\figref{fig::particle_mobility_msd_over_timeinterval_charge_allrealizations}.
For long time intervals~$\tau > \SI{0.1}{\second}$, both patterns (1) and (2) lead to a similar behavior of confined diffusion as obtained for the case of caged particles considered in the previous section\footnote{%
Note however the difference in the diffusive behavior for small time intervals~$\tau < \SI{0.1}{\second}$ that allows to differentiate between purely steric hindrance (cf.~\figref{fig::particle_mobility_msd_over_timeinterval_steric_hindrance_only}) and additional attractive interactions (cf.~\figref{fig::particle_mobility_msd_over_timeinterval_charge_allrealizations}).}%
.
The plateau MSD value thus allows to draw conclusions with respect to the volume enclosed by the particle's initial position and the location(s) of the potential minimum/minima (i.e.,~fiber/charge agglomerations) that the particle has visited.
By looking at the sequence of snapshots over the entire simulation time, it has been verified that those four (highly charged) particles with the smallest MSD plateau values of~$<\Delta r_\text{p}^2> \approx \SI{2e-2}{\micro\meter^2}$ are completely immobilized at one single location in the network (cf.~motion pattern (1)).
In contrast, those medium and highly charged particles with intermediate values for the MSD slopes jump or smoothly transit between local fiber agglomerations in cycles (cf.~motion pattern (2)) such as observed in the example shown in \figref{fig::particle_mobility_cycling}.
Finally, also motion pattern (3) has been identified in the set of individual realizations, as already shown in \figref{fig::particle_mobility_jumps_three_trapped_states_Q8_s3_1} for a particle with high surface charge.
This case can easily be identified among the MSD curves plotted in \figref{fig::particle_mobility_msd_over_timeinterval_charge_allrealizations} as the one with the largest MSD values for long time intervals.

To conclude this section it can be stated that strong attractive forces mostly lead to motion patterns (1) and (2) that effectively immobilize the particles in a confined volume.
However, the motion pattern (3) allows -- at least theoretically -- for a travel of particles over considerable distances by means of a series of successive jumps.
The effectiveness of this transport mechanism depends on the irregularity of the fiber/charge distribution in space and in particular the (relative) height of the potential barriers between the fiber/charge agglomerations.
While the particle in the specific example shown in \figref{fig::particle_mobility_jumps_three_trapped_states_Q8_s3_1} seems to be completely immobilized after two consecutive jumps, one might think of another special design of the fiber/charge distribution with a certain degree of periodicity that could lead to an effective transport of the particle also over large distances and in small time intervals.
Such a directed motion would maximize the mobility of a diffusing particle.

Altogether, the results of this section indicate that the irregularity of the potential landscape exerts a crucial influence on the overall effectiveness of particle immobilization and therefore on the selectivity of hydrogels.
The aforementioned factors that shape the effective potential field explored by the particle are known to vary substantially between the multitude of different biological hydrogels (see e.g.,~\cite{Witten2017}).
Therefore, a systematic parametrization (and if required an extension) of the present computational model with respect to other classes of gels is expected to be a valuable means for studying the species-specific variations of the general behavior and principles observed so far.
In the long run, this might even lead to simulation-based prediction tools enabling a case-specific choice or design of drug delivery vehicles.
Being able to optimize the properties of the carrier to effectively attach the drug to the carrier and yet effectively diffuse through the body would clearly be invaluable in this context.\\

\noindent\textit{Brief discussion of computational aspects.}\\
In total, 60 simulations with at least $\SI{2e4}{}$ time steps each have been conducted and evaluated for the results shown in this section (cf.~\figref{fig::particle_mobility_msd_over_timeinterval_charge_allrealizations} for all realizations).
One simulation typically took 2-4 days if run in parallel on 16 cores%
\footnote{%
AMD Opteron 6128}
on a Linux cluster.
The main drivers for the computational cost are the fiber volume fraction and the presence and strength of electrostatic interactions leading to a complex interplay of repulsive steric and attractive electrostatic forces that require a fine temporal resolution (i.e.,~smaller time steps) and make the nonlinear problem more challenging to solve.

\subsection{The influence of fiber stiffness/compliance}
\label{sec::particle_mobility_results_varying_stiffness}
Up to this point, we haven't discussed the influence of the fiber stiffness, which mainly influences the amount of fiber deformation and -- besides the realistic network geometry -- is the second unique feature of the present computational model.
In the problem setup considered here, fiber deformations originate exclusively from contact and/or electrostatic interactions with the Brownian particle.
Thus, the highest particle charge~$Q_\text{p} = \SI{8e-3}{\femto\coulomb}$ will be considered in this section, because the most frequent and strongest interactions can be expected in this case.
To determine the point where fiber deformations begin to change the results, the value for Young's modulus has been varied systematically starting from the theoretical limit of rigid fibers as outlined already in \secref{sec::particle_mobility_model_setup_fiber_network}.
In this section, the results for rigid fibers will thus be compared to those obtained for a value of~$E=\SI{0.1}{\mega\pascal}$ where the first differences can be observed, and to the ones obtained for a ten times larger value~$E^\ast = 10 \cdot E = \SI{1}{\mega\pascal}$, which are basically identical to the case of rigid fibers.
Note that -- just as in the previous section -- the fiber volume fraction will be kept constant at the medium value~$\bar V_\text{f} \approx 0.4\%$ throughout this section.

\begin{figure}[htb]%
  \centering
  \subfigure[]{
    \includegraphics[width=0.48\textwidth]{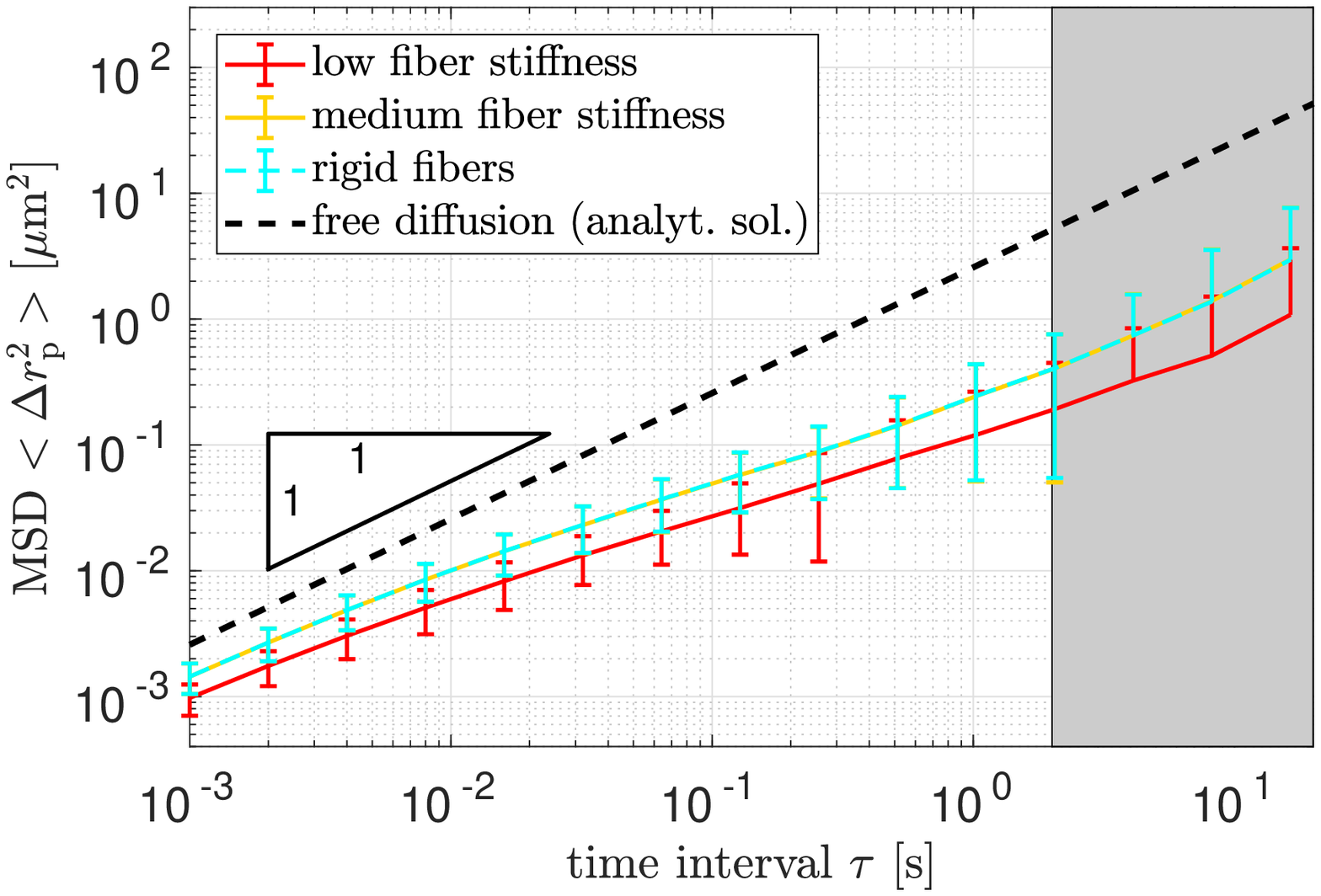}
    \label{fig::particle_mobility_msd_over_timeinterval_fiberstiffness_mean_stddev}
  }
  \subfigure[]{
    \includegraphics[width=0.48\textwidth]{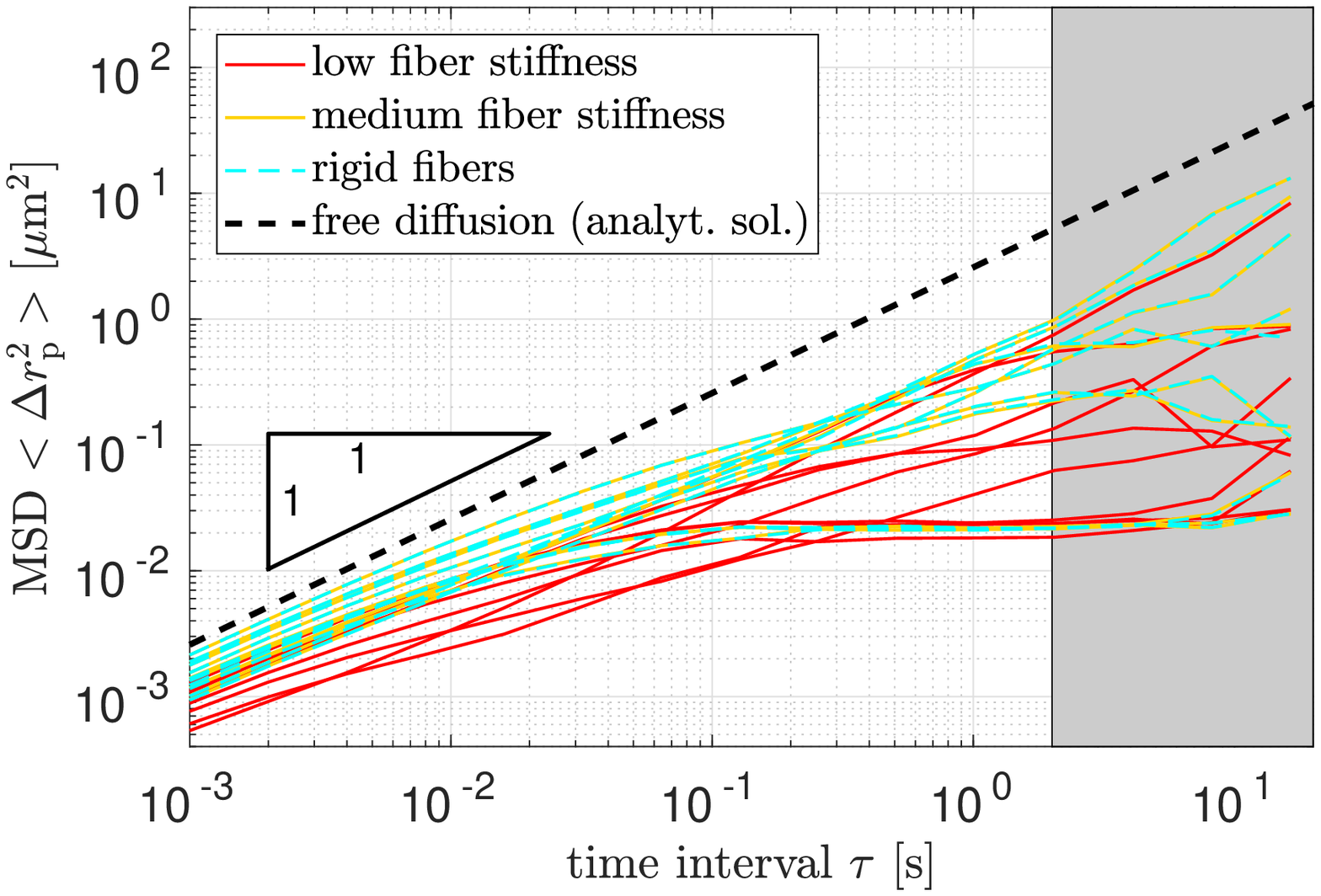}
    \label{fig::particle_mobility_msd_over_timeinterval_fiberstiffness_allrealizations}
  }
  \caption{Analysis of the influence of the fiber stiffness on the (hindrance of) particle mobility (due to attractive electrostatic and repulsive steric interactions with the fiber network).
  Mean squared displacement (MSD) of the particle $<\Delta r_\text{p}^2>$ as a function of the time interval~$\tau$:
  (a) Mean and standard deviation over five random network geometries and two random realizations each for three different levels of the fiber stiffness:
  Low value for Young's modulus~$E=\SI{0.1}{\mega\pascal}$ (red, see also \figref{fig::particle_mobility_msd_over_timeinterval_charge}), medium value for Young's modulus~$E^\ast = \SI{1}{\mega\pascal}$ (yellow), and the limit of rigid fibers (cyan).
  (b) All corresponding individual realizations.
  The fiber volume fraction~$\bar V_\text{f} \approx 0.4\%$ and particle's charge~$Q_\text{p} = \SI{8e-3}{\femto\coulomb}$ are identical for all these realizations.
  The analytical solution for the case of free diffusion is plotted as a reference (black dashed line).
  The gray background indicates the range of time intervals above 10\% of the simulation time, where only few independent samples are available for computing the MSD.
  The bottom of the error bar is hidden for clarity wherever the corresponding value is negative.}
  \label{fig::particle_mobility_msd_over_timeinterval_fiberstiffness}
\end{figure}

\figref{fig::particle_mobility_msd_over_timeinterval_fiberstiffness} compares the MSD curves obtained for the low fiber stiffness resulting from Young's modulus~$E$ (red lines) with the ones obtained for a ten times higher value for Young's modulus~$E^\ast$ (yellow lines) and the ones obtained for the limit of rigid fibers (cyan dashed lines).
Apart from the fiber stiffness, the compared sets of ten realizations each are identical, in particular with respect to the 5 different random network geometries and the 2 different sequences of the random stochastic forces each.\\

\noindent\textbf{The influence of fiber stiffness on particle mobility is insignificant within the range of stiffness values reported for ECM gels.}\\
There is no perceptible difference between the results for the medium fiber stiffness using~$E^\ast$ and those for the limiting case of rigid fibers.
Recall from \secref{sec::particle_mobility_model_setup_fiber_network} that the value~$E^\ast = \SI{1}{\mega\pascal}$ is already the lower limit of the wide range~$E=\SI{1}{\mega\pascal}-\SI{1}{\giga\pascal}$ of values reported for experiments with collagen gels \cite{Jansen2018,Rijt2006}.\\

\noindent\textbf{For more flexible fibers, we observe and speculate about a few competing effects that seem to cause an overall decrease of particle mobility.}\\
First, let us look at the resulting overall effect in terms of the change in ensemble-averaged MSD curves.
For the lowest fiber stiffness~$E = \SI{0.1}{\mega\pascal}$ considered in our simulations, the curves are shifted towards smaller MSD values (see \figref{fig::particle_mobility_msd_over_timeinterval_fiberstiffness}).
The difference is not significant due to the considerable variability already observed in the results of the last section.
However, these results allow to conclude that -- for the given set of parameters -- a value below $E^\ast$ will begin to influence the simulation results.
As an explanation for this behavior, we speculate that the softer fibers lead to an increased adhesive contact area because their shape adapts to the surface shape of the particle and that this behavior in turn traps the oppositely charged particles more effectively.
In addition, the softer fibers will presumably follow the particle's thermal excitation more closely and thus lead to smaller peak values of particle accelerations and interaction forces that act to separate the adhesive contact between fiber(s) and the particle, such that escape events will become less likely.
Both effects might contribute to the observed overall reduction of particle mobility for fibers with a low stiffness.
This reasoning is supported by the previously recognized high importance of trapping and escape events as concluded from the results of the previous \secref{sec::particle_mobility_results_charge}.

\begin{figure}[htb]%
  \centering
  \includegraphics[width=0.48\textwidth]{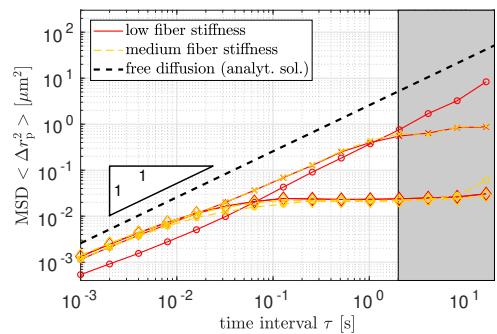}
  \caption{Detailed comparison between low fiber stiffness (red solid lines: $E=\SI{0.1}{\mega\pascal}$) and medium fiber stiffness (yellow dashed lines: $E^\ast = \SI{1}{\mega\pascal}$) for three individual random realizations (crosses, diamonds and circles).
  The plot shows the mean squared displacement (MSD) of the particle $<\Delta r_\text{p}^2>$ as a function of the time interval~$\tau$.
  The analytical solution for the case of free diffusion is plotted as a reference (black dashed line).}
  \label{fig::particle_mobility_msd_over_timeinterval_fiberstiffness_three_realizations}
\end{figure}

\begin{figure}[htb]%
  \centering
  \subfigure[]{
    \includegraphics[width=0.27\textwidth]{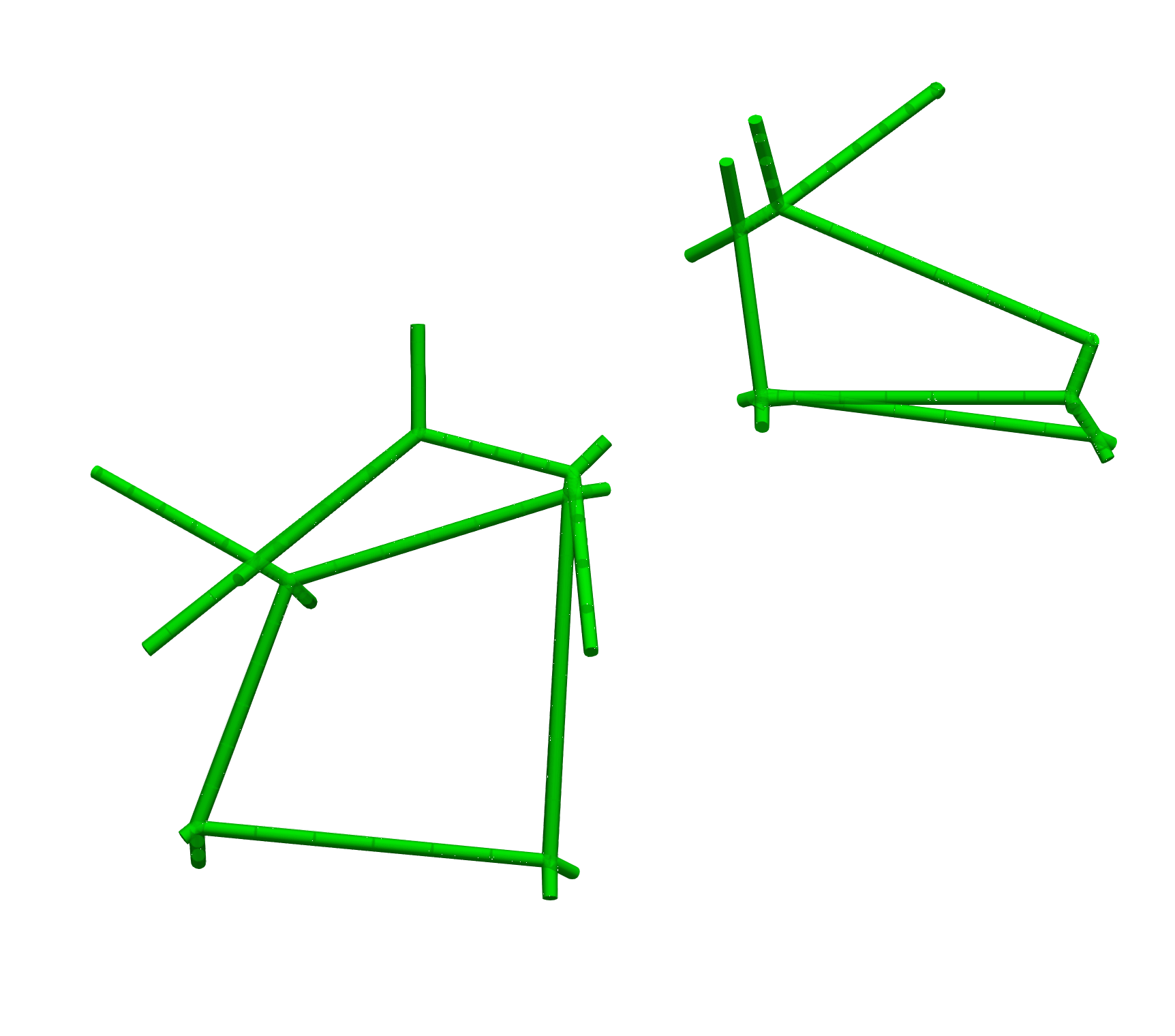}
    \label{fig::particle_mobility_fiberstiffness_displacement_overlay_highstiffness}
  }
  \subfigure[]{
    \includegraphics[width=0.27\textwidth]{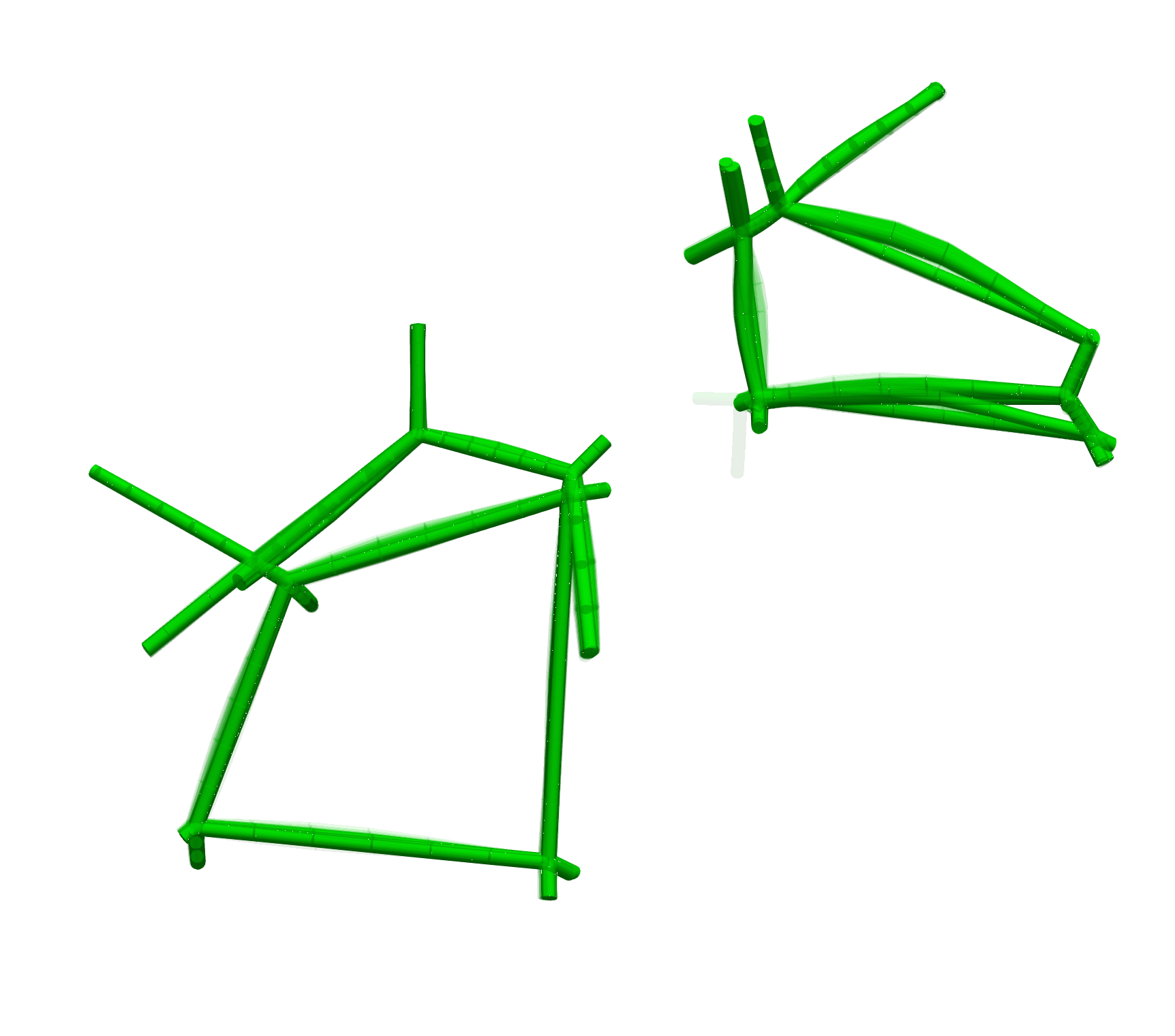}
    \label{fig::particle_mobility_fiberstiffness_displacement_overlay_lowstiffness}
  }
  \subfigure[]{
    \includegraphics[width=0.4\textwidth]{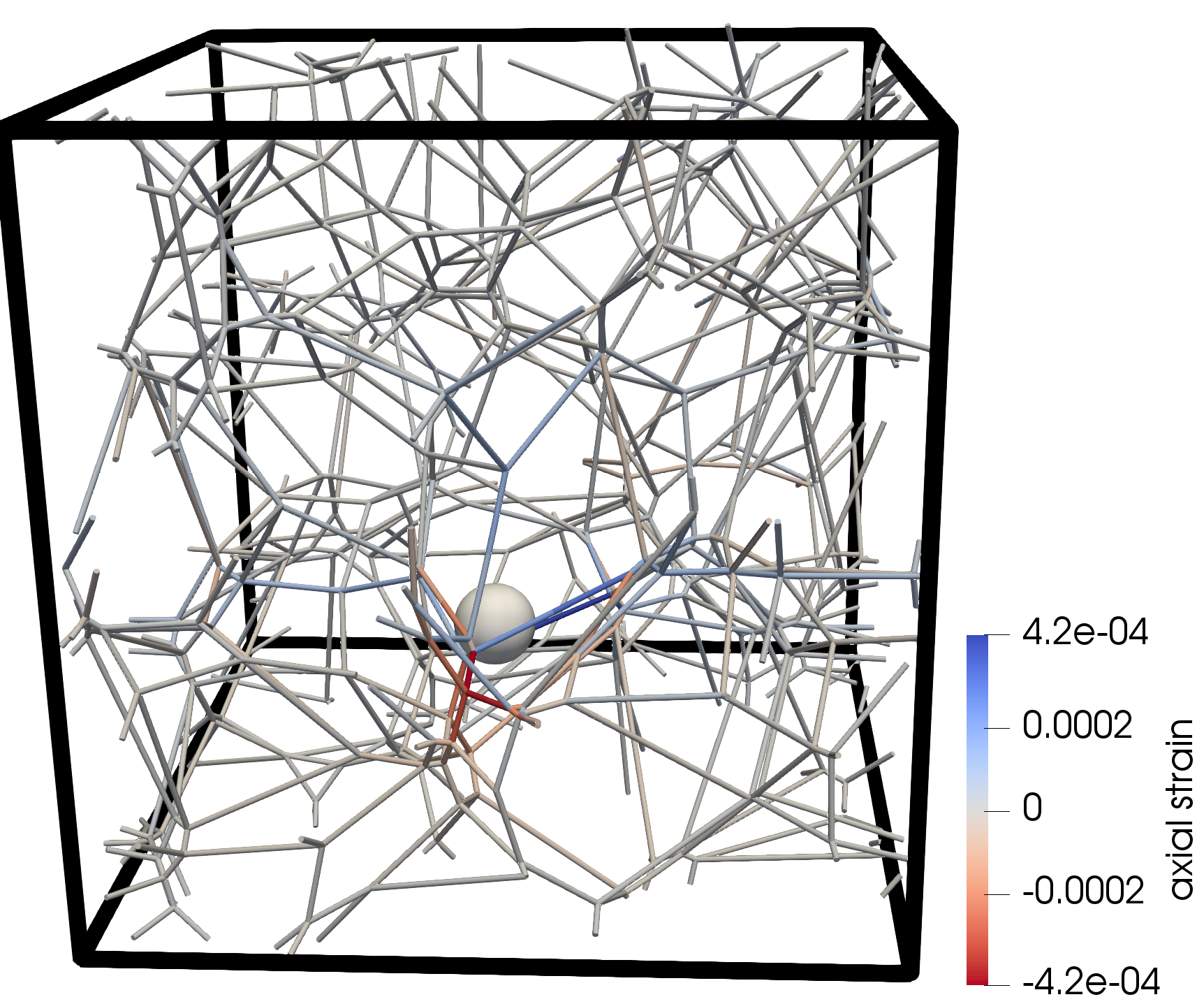}
    \label{fig::particle_mobility_fiberstiffness_strain_cascade}
  }
  \caption{Comparison of the amount of fiber deformations visualized by an overlay of the configurations from all time steps:
  (a) Medium fiber stiffness $E^\ast$ preventing basically any deformations;
  (b) Ten times lower fiber stiffness $E= 0.1 E^\ast$ leading to noticeable, yet small fiber deformations.
  (c) Visualization of the magnitude and distribution of axial strains in the fiber network resulting from the thermal excitation of the particle for a low fiber stiffness $E$.}
  \label{fig::particle_mobility_fiberstiffness_displacement_strains_comparison}
\end{figure}

Generally, there might be more mechanisms how very soft fibers can influence the particle mobility, including some that increase particle mobility.
In an effort to shed some light on this, we look at the direct, pair-wise comparison of a pair of simulations with different fiber stiffness values~$E$ and $E^\ast$ and an otherwise identical setup, i.e., in particular, with identical network geometry, identical initial particle position, and identical sequence of stochastic forces (see \figref{fig::particle_mobility_msd_over_timeinterval_fiberstiffness_three_realizations}).
The first and second pair of realizations with otherwise identical parameters (first pair marked with crosses and second pair marked with diamonds) show nearly identical MSD curves for $E$ (red) and $E^\ast$ (yellow), respectively.
Conversely, the third pair of realizations with otherwise identical parameters (marked with circles) leads to entirely different results for $E$ (red) and $E^\ast$ (yellow).
This leads to the conclusion that small differences in the fiber behavior can trigger entirely different particle trajectories in individual realizations of the stochastic process.
Given the already large variability in the results of individual random realizations for identical fiber stiffness observed in the last section (see the right half of \figref{fig::particle_mobility_msd_over_timeinterval_fiberstiffness_allrealizations}), this sensitivity is not surprising.
The stochastic nature of the problem impedes the direct investigation of differences in particle behavior by means of a pair-wise comparison and underlines the importance of the already analyzed ensemble-averaged results shown in \figref{fig::particle_mobility_msd_over_timeinterval_fiberstiffness_mean_stddev} and discussed above.
However, by taking a closer look at the second pair of realizations and watching the two resulting particle motions as overlaying videos, we can still investigate a quite obvious mechanism that influences the particle mobility providing, in this case, nearly identical MSD curves.
Indeed, one would intuitively argue that stiffer fibers constrain the motion of a particle \textit{more} effectively \textit{as long as it remains in the state of adhesive contact with the fiber}, because the softer fibers deform due to the thermal forces acting on the trapped particle.
This situation is shown in Figures~\ref{fig::particle_mobility_fiberstiffness_displacement_overlay_lowstiffness} and \ref{fig::particle_mobility_fiberstiffness_displacement_overlay_highstiffness}, which show an overlay of fiber configurations over all time steps for these two realizations.
The two corresponding MSD curves in \figref{fig::particle_mobility_msd_over_timeinterval_fiberstiffness_three_realizations} indeed confirm that the softer fibers (red diamonds) allow for slightly higher MSD values than the stiffer fibers (yellow diamonds) on time intervals~$\tau \geq \SI{0.1}{\second}$.
However, these MSD curves also show that this effect is rather negligible in terms of overall mobility of the particles.
This is also supported by recognizing the small magnitude and localized extent of axial strains in the network (at an exemplarily chosen point in time) in \figref{fig::particle_mobility_fiberstiffness_strain_cascade}.\\

To conclude this first brief study of how deformations of semiflexible fibers influence the mobility of oppositely charged particles, it can thus be stated that our results indicate a -- rather counter-intuitive -- overall \textit{decrease} in particle mobility if compared to (almost) rigid fibers.
We suggest that this is the result of a decreased probability for escape events due to a) an increase in the adhesive contact area between particle and fiber and b) a higher ability of the fibers to follow the thermal excitation of a trapped particle, thus reducing the peak values of particle accelerations and interaction forces that act to break the adhesive binding.
It is important to note, however, that this influence of the fiber stiffness on the particle mobility is negligible in the range of stiffness values that have been reported for the collagen I fibers prevailing in ECM gels.
Within this range, the fibers do not show any noticeable deformations for the scenario considered here and the associated influence on the particle trajectory thus vanishes.
Nevertheless, the influence of fiber deformations is likely to be observed for other types of (biological) hydrogels with very thin or soft and therefore more flexible filaments such as mucin or F-actin, and maybe even for ECM gels as a result of dysregulated fiber stiffness.
A more detailed investigation of this aspect, both in simulations and experiments, is thus considered a promising avenue of future research.

\section{Conclusions and outlook}
\label{sec::particle_mobility_conclusion_outlook}
This article presents the first computational study of the diffusive mobility of particles in hydrogels with a realistic fiber network model.
It proposes a novel computational approach based on, most notably, the modeling of individual, deformable fibers via the beam theory, the Voronoi tessellation of the periodic simulation box to obtain random, irregular network geometries, and the beam-sphere interaction model for contact and electrostatic interactions.
Following the validation of the model and the study of repulsive steric interactions only, the particularly important effect of additional attractive electrostatic forces has been investigated.
Finally, we have studied the role of fiber deformations in the latter case by means of additional computational experiments.

In the case of only repulsive steric interactions, it is found that the hindrance of the particle mobility is insignificant as long as the mesh sizes of the fiber network are larger than the particle diameter.
If, however, the mesh sizes are in the order of the particle diameter, the particle is effectively caged in a polygonal fiber hull of random shape and size and shows a behavior known as confined diffusion and is characterized by a plateau in the mean squared displacement (MSD) curve for long time intervals.
Within the given problem setup and the focus on relatively small particles, these are expected results validating the novel approach.
However, this effect of steric hindrance will become highly relevant if the effective transport of relatively large particles as observed in experiments \cite{Lai2007} is considered.
In this context, including the dynamics of the fiber network such as its self-assembly driven by the Brownian motion and transient reorganization of cross-links (as demonstrated for the directly compatible computational model applied in \cite{Cyron2013phasediagram,Cyron2013a}) might be an important model component as suggested by recent findings in the context of the nuclear pore complex \cite{Goodrich2018,Colwell2010} or the dynamic secretion and shedding of mucus layers \cite{Marczynski2018}.

Turning to the effect of additional electrostatic interactions between the fibers and the oppositely charged particle, the prevailing notion that the degree of hindrance on average increases with the strength of attraction has been confirmed by the numerical experiments with five different network geometries and two random realizations each.
Moreover, an increased variability of the particle's mean squared displacement values and slopes in the regime of long time intervals has been observed and excellently agrees with previous experimental results \cite{Arends2013}.
A detailed look at the 3D particle trajectories within the fiber network provides a first direct proof for the existence of distinct motion patterns of the particles, which explains the variability in the MSD curves.
As hypothesized in the previous work \cite{Arends2013}, the particles stick to oppositely charged fiber/charge aggregations experiencing more or less strong trapping and eventually escape due to the ongoing thermal excitation, only to be quickly attracted to another fiber/charge aggregation.
While some particles remain completely immobilized at one and the same location for the entire 20s of simulation time, others smoothly or rapidly cycle between two local minima in the potential landscape.
Both of these motion patterns lead to a behavior on longer time intervals that is very similar to the confined diffusion for caged, uncharged particles as described above.
However, the diffusive mobility on short time intervals is significantly reduced as well due to sticking to the fibers.
The third motion pattern observed is the one of several successive jumps that -- at least theoretically -- could serve as a transport mechanism also over longer distances if the potential landscape is formed accordingly and e.g.,~shows some degree of periodicity and directional preference.

Altogether, these findings indicate that the precise shape of the effective 3D potential field explored by the particle has a crucial influence on its mobility.
In view of the broad variety of biopolymer hydrogels with diverse chemical compositions and biophysical properties, the current computational model could thus be leveraged to study the individual selective filtering behavior for a large number of particle-hydrogel property combinations.
Based on the recognized importance of the precise fiber/charge distribution in the system, two points seem to be of particular importance to achieve a case-specific, highly accurate and reliable prediction.
First, the inhomogeneous charge distribution along the fiber should be both determined (e.g.,~by experimentally analyzing the molecular architecture) and applied in the model.
Second, the specific composition and geometry of fiber networks should be determined (e.g.,~by processing electron microscopy images) and applied in the model.
Also the inclusion of the dynamic self-assembly and reorganization of networks mentioned above would be worth considering in this respect.

As a last particular aspect investigated in this study, fiber deformations have been found to be negligible within the range of realistic values for the stiffness of collagen I fibers prevailing in ECM gels.
To be more precise, varying the value for the Young's modulus over the broad range of reported values for the considered ECM gels has led to identical results as obtained for the theoretical limit of rigid fibers.
If, however, more flexible fibers are considered, our simulation results indicate an overall decrease of particle mobility if compared to (almost) rigid fibers  -- an outcome that is rather counter-intuitive.
We suggest that this is the result of a decreased probability for escape events due to a) an increase in the adhesive contact area between particle and fiber and b) a higher ability of the fibers to follow the thermal excitation of a trapped particle, thus reducing the peak values of particle accelerations and interaction forces that act to break the adhesive binding.
In real systems, these trends might be observed for other types of (biological) hydrogels with very thin or soft and therefore more flexible filaments such as mucin or F-actin, and maybe even for ECM gels as a result of dysregulated fiber stiffness.
A more detailed investigation of this aspect, both in simulations and experiments, is thus considered a promising avenue of future research.

In addition to the presented simulation results and the gained insights, this study provides an extensive proof of concept for the application of the novel computational model.
As outlined above, in the short to medium term many important findings especially for various particle/hydrogel-specific behaviors and mechanisms can be expected from applying different parametrizations and, additionally, from integrating the suggested model extensions.
In the long term, further validation and advances of the present computational model toward a truly predictive tool could ultimately lead to a case- and patient-specific choice or even design of pharmaceuticals and also to a case- and patient-specific assessment of infection risk.

\vskip6pt

\enlargethispage{20pt}

\ethics{This article does not contain any studies with human or animal subjects.}

\dataccess{All required information to reproduce the results of the numerical experiments is provided or referenced in this article. The source code is part of a wider code base~\cite{BACI2020} that will be made available in the future. The data is provided as electronic supplementary material.}

\aucontribute{All authors contributed to writing the manuscript. WAW, OL and MJG designed the research. MJG, CM and JFE developed the numerical model and methods. JK and MJG performed the numerical experiments and analyzed the data.}

\competing{All authors declare that there is no conflict of interest.}

\ack{We thank Kei W.~M\"uller for useful discussions.}

\vskip2pc

\bibliographystyle{RS}
\bibliography{library_particle_mobility_in_hydrogels.bib}

\begin{thebibliography}{99}

\bibitem{Witten2017}
Witten J, Ribbeck K. 2017  {The particle in the spider's web: transport through
  biological hydrogels}. {\em Nanoscale} \textbf{9}, 8080--8095.

\bibitem{Bray1984}
Bray J, Robinson GB, Byrne J. 1984  {Influence of charge on filtration across
  renal basement membrane films in vitro}. {\em Kidney International}
  \textbf{25}, 527--533.

\bibitem{Dowd1999}
Dowd CJ, Cooney CL, Nugent MA. 1999  {Heparan Sulfate Mediates bFGF Transport
  through Basement Membrane by Diffusion with Rapid Reversible Binding}. {\em
  Journal of Biological Chemistry} \textbf{274}, 5236--5244.

\bibitem{Dellian2000}
Dellian M, Yuan F, Trubetskoy VS, Torchilin VP, Jain RK. 2000  {Vascular
  permeability in a human tumour xenograft: molecular charge dependence}. {\em
  British Journal of Cancer} \textbf{82}, 1513--1518.

\bibitem{Olmsted2001}
Olmsted SS, Padgett JL, Yudin AI, Whaley KJ, Moench TR, Cone RA. 2001
  {Diffusion of Macromolecules and Virus-Like Particles in Human Cervical
  Mucus}. {\em Biophysical Journal} \textbf{81}, 1930--1937.

\bibitem{Lai2007}
Lai SK, O'Hanlon DE, Harrold S, Man ST, Wang YY, Cone R, Hanes J. 2007  {Rapid
  transport of large polymeric nanoparticles in fresh undiluted human mucus}.
  {\em Proceedings of the National Academy of Sciences} \textbf{104}, 1482 --
  1487.

\bibitem{Lieleg2009}
Lieleg O, Baumg{\"{a}}rtel RM, Bausch AR. 2009  {Selective filtering of
  particles by the extracellular matrix: an electrostatic bandpass}. {\em
  Biophysical Journal} \textbf{97}, 1569--1577.

\bibitem{Lieleg2010}
Lieleg O, Vladescu I, Ribbeck K. 2010  {Characterization of Particle
  Translocation through Mucin Hydrogels}. {\em Biophysical Journal}
  \textbf{98}, 1782--1789.

\bibitem{Colwell2010}
Colwell LJ, Brenner MP, Ribbeck K. 2010  {Charge as a Selection Criterion for
  Translocation through the Nuclear Pore Complex}. {\em PLOS Computational
  Biology} \textbf{6}, e1000747.

\bibitem{Schuster2013}
Schuster BS, Suk JS, Woodworth GF, Hanes J. 2013  {Nanoparticle diffusion in
  respiratory mucus from humans without lung disease}. {\em Biomaterials}
  \textbf{34}, 3439--3446.

\bibitem{Xu2013}
Xu Q, Boylan NJ, Suk JS, Wang YY, Nance EA, Yang JC, McDonnell PJ, Cone RA, Duh
  EJ, Hanes J. 2013  {Nanoparticle diffusion in, and microrheology of, the
  bovine vitreous ex vivo}. {\em Journal of Controlled Release} \textbf{167},
  76--84.

\bibitem{Arends2013}
Arends F, Baumg{\"{a}}rtel R, Lieleg O. 2013  {Ion-Specific Effects Modulate
  the Diffusive Mobility of Colloids in an Extracellular Matrix Gel}. {\em
  Langmuir} \textbf{29}, 15965--15973.

\bibitem{Zhang2015}
Zhang X, Hansing J, Netz RR, DeRouchey JE. 2015  {Particle Transport through
  Hydrogels Is Charge Asymmetric}. {\em Biophysical Journal} \textbf{108},
  530--539.

\bibitem{Kasdorf2015}
K{\"{a}}sdorf BT, Arends F, Lieleg O. 2015  {Diffusion Regulation in the
  Vitreous Humor}. {\em Biophysical Journal} \textbf{109}, 2171--2181.

\bibitem{Abdulkarim2015}
Abdulkarim M, Agull{\'{o}} N, Cattoz B, Griffiths P, Bernkop-Schn{\"{u}}rch A,
  Borros SG, Gumbleton M. 2015  {Nanoparticle diffusion within intestinal
  mucus: Three-dimensional response analysis dissecting the impact of particle
  surface charge, size and heterogeneity across polyelectrolyte, pegylated and
  viral particles}. {\em European Journal of Pharmaceutics and
  Biopharmaceutics} \textbf{97}, 230--238.

\bibitem{Arends2017}
Arends F, Chaudhary H, Janmey P, Claessens MMAE, Lieleg O. 2017  {Lipid Head
  Group Charge and Fatty Acid Configuration Dictate Liposome Mobility in
  Neurofilament Networks}. {\em Macromolecular Bioscience} \textbf{17},
  1600229.

\bibitem{Johansson1993}
Johansson L, L{\"{o}}froth JE. 1993  {Diffusion and interaction in gels and
  solutions. 4. Hard sphere Brownian dynamics simulations}. {\em The Journal of
  Chemical Physics} \textbf{98}, 7471--7479.

\bibitem{Saxton1994}
Saxton M. 1994  {Anomalous diffusion due to obstacles: a Monte Carlo study}.
  {\em Biophysical Journal} \textbf{66}, 394--401.

\bibitem{Netz1997}
Netz PA, Dorfm{\"{u}}ller T. 1997  {Computer simulation studies of diffusion in
  gels: Model structures}. {\em The Journal of Chemical Physics} \textbf{107},
  9221--9233.

\bibitem{Pei2009}
Pei H, Allison S, Haynes BMH, Augustin D. 2009  {Brownian Dynamics Simulation
  of the Diffusion of Rods and Wormlike Chains in a Gel Modeled as a Cubic
  Lattice: Application to DNA}. {\em The Journal of Physical Chemistry B}
  \textbf{113}, 2564--2571.

\bibitem{Stylianopoulos2010fiberorientation}
Stylianopoulos T, Diop-Frimpong B, Munn LL, Jain RK. 2010  {Diffusion
  Anisotropy in Collagen Gels and Tumors: The Effect of Fiber Network
  Orientation}. {\em Biophysical Journal} \textbf{99}, 3119--3128.

\bibitem{Kamerlin2016}
Kamerlin N, Elvingson C. 2016  {Tracer diffusion in a polymer gel: Simulations
  of static and dynamic 3D networks using spherical boundary conditions}. {\em
  Journal of Physics Condensed Matter} \textbf{28}.

\bibitem{Saxton1996}
Saxton M. 1996  {Anomalous diffusion due to binding: a Monte Carlo study}. {\em
  Biophysical Journal} \textbf{70}, 1250--1262.

\bibitem{Zhou2009}
Zhou H, Chen SB. 2009  {Brownian dynamics simulation of tracer diffusion in a
  cross-linked network}. {\em Physical Review E} \textbf{79}, 21801.

\bibitem{Stylianopoulos2010electrostatics}
Stylianopoulos T, Poh MZ, Insin N, Bawendi MG, Fukumura D, Munn LLL, Jain RK.
  2010  {Diffusion of Particles in the Extracellular Matrix: The Effect of
  Repulsive Electrostatic Interactions}. {\em Biophysical Journal} \textbf{99},
  1342--1349.

\bibitem{Miyata2012}
Miyata T. 2012  {Brownian Dynamics Simulation of Self-Diffusion of Ionic Large
  Solute Molecule in Modeled Polyelectrolyte Gel}. {\em Journal of the Physical
  Society of Japan} \textbf{81}, SA010.

\bibitem{Hansing2016}
Hansing J, Ciemer C, Kim WK, Zhang X, DeRouchey JE, Netz RR. 2016
  {Nanoparticle filtering in charged hydrogels: Effects of particle size,
  charge asymmetry and salt concentration}. {\em European Physical Journal E}
  \textbf{39}, 53.

\bibitem{Hansing2018}
Hansing J, Netz RR. 2018  {Particle Trapping Mechanisms Are Different in
  Spatially Ordered and Disordered Interacting Gels}. {\em Biophysical Journal}
  \textbf{114}, 2653--2664.

\bibitem{Humphries2017}
Humphries DL, Grogan JA, Gaffney EA. 2017  {Mechanical Cell-Cell Communication
  in Fibrous Networks: The Importance of Network Geometry}. {\em Bulletin of
  Mathematical Biology} \textbf{79}, 498--524.

\bibitem{Cyron2010}
Cyron CJ, Wall WA. 2010  {Consistent finite-element approach to Brownian
  polymer dynamics with anisotropic friction}. {\em Physical Review E}
  \textbf{82}, 66705.

\bibitem{Slepukhin2019}
Slepukhin VM, Grill MJ, M{\"{u}}ller KW, Wall WA, Levine AJ. 2019
  {Conformation of a semiflexible filament in a quenched random potential}.
  {\em Physical Review E} \textbf{99}, 042501.

\bibitem{Cyron2013phasediagram}
Cyron CJ, M{\"{u}}ller KW, Schmoller KM, Bausch AR, Wall WA, Bruinsma RF. 2013a
   {Equilibrium phase diagram of semi-flexible polymer networks with linkers}.
  {\em Europhysics Letters} \textbf{102}, 38003.

\bibitem{Cyron2013a}
Cyron CJ, M{\"{u}}ller KW, Bausch AR, Wall WA. 2013b  {Micromechanical
  simulations of biopolymer networks with finite elements}. {\em Journal of
  Computational Physics} \textbf{244}, 236--251.

\bibitem{Mueller2014rheology}
M{\"{u}}ller KW, Bruinsma RF, Lieleg O, Bausch AR, Wall WA, Levine AJ. 2014
  {Rheology of Semiflexible Bundle Networks with Transient Linkers}. {\em
  Physical Review Letters} \textbf{112}, 238102.

\bibitem{Goodrich2018}
Goodrich CP, Brenner MP, Ribbeck K. 2018  {Enhanced diffusion by binding to the
  crosslinks of a polymer gel}. {\em Nature Communications} \textbf{9}, 4348.

\bibitem{Marczynski2018}
Marczynski M, K{\"{a}}sdorf BT, Altaner B, Wenzler A, Gerland U, Lieleg O. 2018
   {Transient binding promotes molecule penetration into mucin hydrogels by
  enhancing molecular partitioning}. {\em Biomaterials Science} \textbf{6},
  3373--3387.

\bibitem{Meier2017c}
Meier C, Popp A, Wall WA. 2019  {Geometrically Exact Finite Element
  Formulations for Slender Beams: Kirchhoff-Love Theory Versus Simo-Reissner
  Theory}. {\em Archives of Computational Methods in Engineering} \textbf{26},
  163--243.

\bibitem{Meier2017b}
Meier C, Grill MJ, Wall WA, Popp A. 2018  {Geometrically exact beam elements
  and smooth contact schemes for the modeling of fiber-based materials and
  structures}. {\em International Journal of Solids and Structures}
  \textbf{154}, 124--146.

\bibitem{GrillSSIP}
Grill MJ, Wall WA, Meier C. 2020  {A computational model for molecular
  interactions between curved slender fibers undergoing large 3D deformations
  with a focus on electrostatic, van der Waals, and repulsive steric forces}.
  {\em International Journal for Numerical Methods in Engineering}
  \textbf{121}, 2285--2330.

\bibitem{Arends2015PLOS}
Arends F, Nowald C, Pflieger K, Boettcher K, Zahler S, Lieleg O. 2015  {The
  biophysical properties of basal lamina gels depend on the biochemical
  composition of the gel}. {\em PLoS ONE} \textbf{10}, e0118090.

\bibitem{BACI2020}
{BACI: A Comprehensive Multi-Physics Simulation Framework}. 2020
  {https://baci.pages.gitlab.lrz.de/website}. .

\bibitem{Rycroft2009}
Rycroft C. 2009  {Voro++: a three-dimensional Voronoi cell library in C++}.
  Technical report United States.

\bibitem{Blender2_80}
{Blender Foundation}. 2019  {Blender 2.80}. .

\bibitem{Yurchenco1987}
Yurchenco PD, Ruben GC. 1987  {Basement membrane structure in situ: evidence
  for lateral associations in the type IV collagen network.}. {\em The Journal
  of Cell Biology} \textbf{105}, 2559--2568.

\bibitem{reissner1981}
Reissner E. 1981  {On finite deformations of space-curved beams}. {\em
  Zeitschrift f{\"{u}}r Angewandte Mathematik und Physik (ZAMP)} \textbf{32},
  734--744.

\bibitem{simo1985}
Simo JC. 1985  {A Finite Strain Beam Formulation. The Three-Dimensional Dynamic
  Problem. Part I}. {\em Computer Methods in Applied Mechanics and Engineering}
  \textbf{49}, 55--70.

\bibitem{simo1986}
Simo JC, {Vu Quoc} L. 1986  {A Three Dimensional Finite Strain Rod Model Part
  II: Computational Aspects}. {\em Computer Methods in Applied Mechanics and
  Engineering} \textbf{58}, 79--116.

\bibitem{Jansen2018}
Jansen KA, Licup AJ, Sharma A, Rens R, MacKintosh FC, Koenderink GH. 2018  {The
  Role of Network Architecture in Collagen Mechanics}. {\em Biophysical
  Journal} \textbf{114}, 2665--2678.

\bibitem{Rijt2006}
van~der Rijt JAJ, van~der Werf KO, Bennink ML, Dijkstra PJ, Feijen J. 2006
  {Micromechanical Testing of Individual Collagen Fibrils}. {\em Macromolecular
  Bioscience} \textbf{6}, 697--702.

\bibitem{jelenic1999}
Jeleni{\'{c}} G, Crisfield MA. 1999  {Geometrically exact 3D beam theory:
  implementation of a strain-invariant finite element for statics and
  dynamics}. {\em Computer Methods in Applied Mechanics and Engineering}
  \textbf{171}, 141--171.

\bibitem{crisfield1999}
Crisfield MA, Jeleni{\'{c}} G. 1999  {Objectivity of strain measures in the
  geometrically exact three-dimensional beam theory and its finite-element
  implementation}. {\em Proceedings of the Royal Society of London. Series A:
  Mathematical, Physical and Engineering Sciences} \textbf{455}, 1125--1147.

\bibitem{GrillPeelingPulloff}
Grill MJ, Meier C, Wall WA. 2019  {Investigation of the peeling and pull-off
  behavior of adhesive elastic fibers via a novel computational beam
  interaction model}. {\em The Journal of Adhesion,
  https://doi.org/10.1080/00218464.2019.1699795}.

\bibitem{Doi1988}
Doi M, Edwards SF. 1988 {\em {The theory of polymer dynamics}}.
Oxford university press.

\bibitem{meier2016}
Meier C, Popp A, Wall WA. 2016  {A finite element approach for the line-to-line
  contact interaction of thin beams with arbitrary orientation}. {\em Computer
  Methods in Applied Mechanics and Engineering} \textbf{308}, 377--413.

\bibitem{Kusumi1993}
Kusumi A, Sako Y, Yamamoto M. 1993  {Confined lateral diffusion of membrane
  receptors as studied by single particle tracking (nanovid microscopy).
  Effects of calcium-induced differentiation in cultured epithelial cells}.
  {\em Biophysical Journal} \textbf{65}, 2021--2040.

\end{thebibliography}

\end{document}